\documentclass[twocolumn,pra,aps,superscriptaddress]{revtex4}
\usepackage{graphicx}
\usepackage{dcolumn}
\usepackage{bm}
\usepackage{bbm}
\usepackage[utf8]{inputenc}
\usepackage{amssymb}
\usepackage{amsmath}
\usepackage{bbold}
\usepackage{pstricks}
\usepackage{slashed}
\usepackage{braket}
\usepackage{hyperref}
\usepackage{natbib}
\usepackage{float}
\usepackage{verbatim}
\usepackage{siunitx}
\usepackage{lipsum}
\usepackage{slashed}
\DeclareSIUnit\gauss{G}

\definecolor{mygreen}{rgb}{0,0.5,0}
\definecolor{myblue}{rgb}{0,0,0.75}
\definecolor{mymagenta}{cmyk}{0,1,0,0.12}
\definecolor{mygray}{rgb}{0.5,0.5,0.5}

\usepackage{umoline}

\begin{document}

\title{Quantum simulation of lattice gauge theories using Wilson fermions}

\author{T. V. Zache}
\email[]{zache@thphys.uni-heidelberg.de}
\affiliation{Heidelberg University, Institut f\"{u}r Theoretische Physik, Philosophenweg 16,
D-69120 Heidelberg}

\author{F. Hebenstreit}
\affiliation{University of Bern, Albert Einstein Center for Fundamental Physics, Institute for Theoretical Physics, Sidlerstrasse 5, CH-3012 Bern}

\author{F. Jendrzejewski }
\affiliation{Heidelberg University, Kirchhoff-Institute f\"{u}r Physik,
Im Neuenheimer Feld 227, D-69120 Heidelberg}

\author{M. K. Oberthaler}
\affiliation{Heidelberg University, Kirchhoff-Institute f\"{u}r Physik,
Im Neuenheimer Feld 227, D-69120 Heidelberg}

\author{J. Berges}
\affiliation{Heidelberg University, Institut f\"{u}r Theoretische Physik, Philosophenweg 16,
D-69120 Heidelberg}

\author{P. Hauke}
\affiliation{Heidelberg University, Institut f\"{u}r Theoretische Physik, Philosophenweg 16,
D-69120 Heidelberg}
\affiliation{Heidelberg University, Kirchhoff-Institute f\"{u}r Physik,
Im Neuenheimer Feld 227, D-69120 Heidelberg}

\date{\today}

\begin{abstract}

Quantum simulators have the exciting prospect of giving access to real-time dynamics of lattice gauge theories, in particular in regimes that are difficult to compute on classical computers. Future progress towards scalable quantum simulation of lattice gauge theories, however, hinges crucially on the efficient use of experimental resources. As we argue in this work, due to the fundamental non-uniqueness of discretizing the relativistic Dirac Hamiltonian, the lattice representation of gauge theories allows for an optimization that up to now has been left unexplored. We exemplify our discussion with lattice quantum electrodynamics in two-dimensional space-time, where we show that the formulation through Wilson fermions provides several advantages over the previously considered staggered fermions. Notably, it enables a strongly simplified optical lattice setup and it reduces the number of degrees of freedom required to simulate dynamical gauge fields. Exploiting the optimal representation, we propose an experiment based on a mixture of ultracold atoms trapped in a tilted optical lattice. Using numerical benchmark simulations, we demonstrate that a state-of-the-art quantum simulator may access the Schwinger mechanism and map out its non-perturbative onset.
\end{abstract}


\maketitle

\section{Introduction}

Recent years have seen considerable progress towards quantum simulations of gauge theories that describe the fundamental interplay of fermionic matter with dynamical gauge fields \cite{wiese2013ultracold,dalmonte2016lattice,zohar2015quantum}. 
By building on dramatic experimental advances, proposals have been presented for optical lattices \cite{banerjee2012atomic,szpak2012optical,zohar2013simulating,tagliacozzo2013optical}, ion chains \cite{hauke2013quantum,yang2016analog}, and superconducting qubits \cite{marcos2013superconducting,mezzacapo2015non}. 
These proposals aim to implement the sophisticated framework of lattice gauge theory in table-top experiments, 
which has enabled numerical calculations for theories such as quantum electrodynamics (QED) or quantum chromodynamics (QCD) \cite{rothe2005lattice}. 
While the sign problem restricts these classical computer calculations mainly to static properties at rather low fermion densities (except for specific limiting cases or under various degrees of approximation), quantum simulators do not suffer from such limitations \cite{georgescu2014quantum}. 
This prospect has motivated a first proof-of-principle implementation on a trapped-ion quantum computer~\cite{martinez2016real}, which has observed the out-of-equilibrium dynamics of a gauge theory, but restricted to an Abelian symmetry, one spatial dimension, rather short times, and few qubits.  

Further progress hinges crucially on efficient implementations, such that present state-of-the-art experimental resources become sufficient to quantum simulate relevant physical processes, such as Schwinger pair production of fermions and anti-fermions in the presence of strong electric fields~\cite{sauter1931verhalten,schwinger1951gauge} or string breaking due to
confinement~\cite{philipsen1998string,bali2005observation,hebenstreit2013real}.
In view of finding optimal implementations, the Nielsen-Ninomiya  no-go-theorem~\cite{nielsen1981absence} becomes particularly important: it states that it is not possible to discretize relativistic fermions  while retaining the relevant symmetries of the continuum theory.
Being forced to make a choice between discretizations with different  symmetry properties, one should be guided by the requirements of the task at hand, e.g, by conceptual advantages, numerical efficiency, or --- as in our work --- ease of experimental implementation. So far, however, most proposals for the engineering of quantum simulators for lattice gauge theories employ one specific discretization procedure via the so-called staggered fermion formulation~\cite{susskind1977lattice}.

\begin{figure}
\includegraphics[scale=0.6]{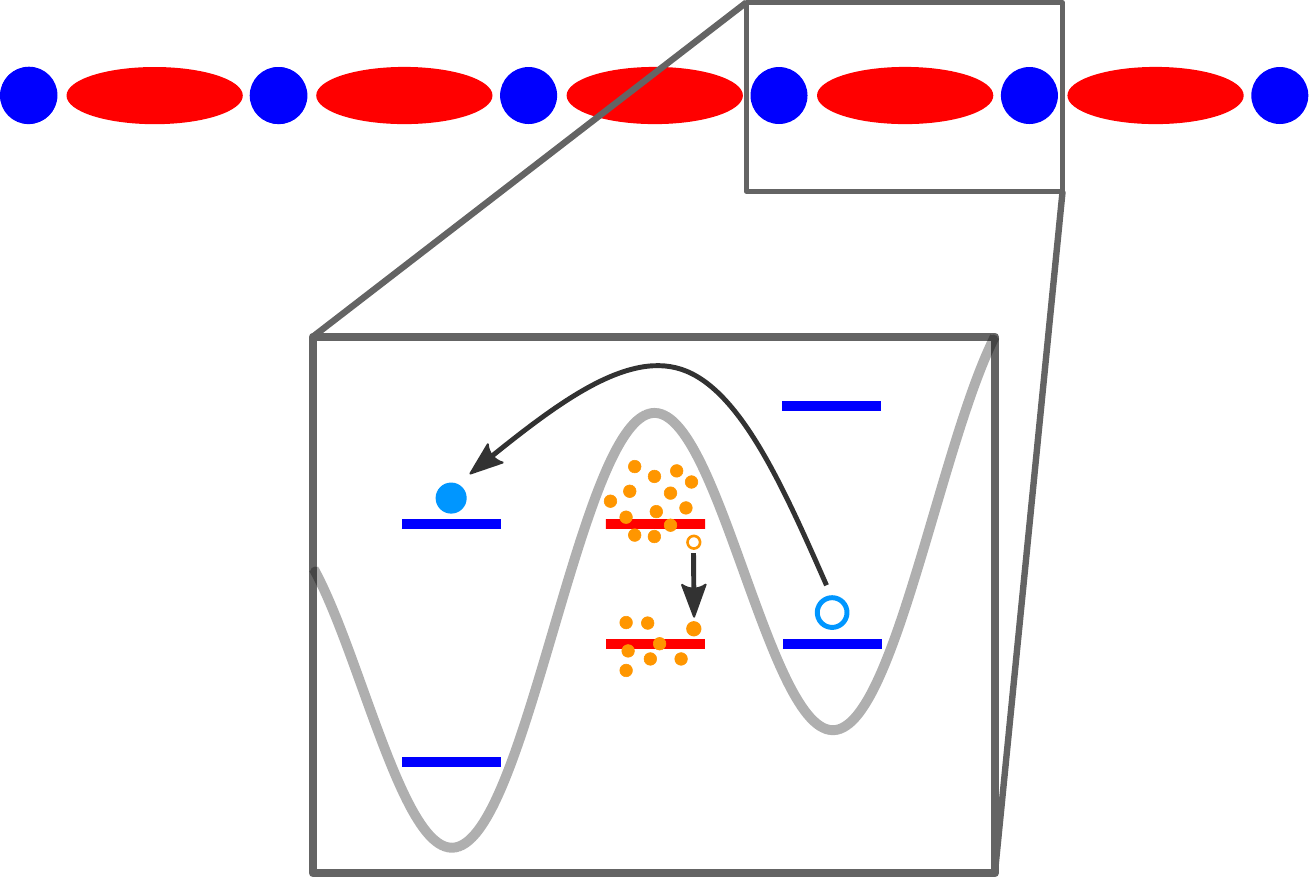}
\caption{\label{sketch}Sketch of the proposed implementation of lattice QED in one spatial dimension. Fermions trapped on the lattice sites (blue circles) are coupled via a correlated interaction with the Bose condensates residing on the links (red ellipses). The zoom schematically shows how the gauge-invariant coupling can be realized via spin-changing collisions between the fermions (blue) and bosons (orange) in a tilted optical lattice. This process involves two internal states per species as indicated by the blue and red bars.}
\end{figure}

In this work, we propose the use of an alternative discretization based on Wilson fermions \cite{wilson1977quarks}. 
As we discuss below, Wilson fermions have conceptual advantages over staggered fermions when going to higher dimensions.  
Moreover, we show that Wilson fermions can provide a very efficient framework for the experimental implementation of gauge theories using ultracold atoms in optical lattices. 
As an example, we discuss QED in $1+1$ space-time dimensions implemented with a two-species mixture. Radiofrequency-dressed ${}^6 \text{Li}$ atoms act as the fermionic matter, while small condensates of bosonic ${}^{23} \text{Na}$ atoms represent the dynamical gauge fields. 
Inter-species spin-changing collisions generate the dynamics of an interacting gauge theory, where local gauge invariance is ensured by angular momentum and energy conservation, see Fig.~\ref{sketch}.  
Strikingly, the use of Wilson fermions enables an implementation through tilted optical lattices, instead of the more involved superlattices employed for staggered fermions \cite{banerjee2012atomic}.
We benchmark our proposal by a theoretical analysis, and show that the non-perturbative onset of Schwinger pair production may be observed in realistic experimental settings.

Wilson fermions have been considered previously in a cold-atom context for the quantum simulation of topological insulators \cite{bermudez2010wilson,mazza2012optical,kuno2018generalized}. In contrast, we are interested here in the full, interacting quantum theory with dynamical gauge fields. 

The paper is organized as follows. 
In Sec.~\ref{sec2}, we present the lattice Hamiltonian of Wilson fermions. We show that an optimal choice of parameters significantly simplifies the resulting setup, and we compare it to the staggered-fermion formulation. 
In Sec.~\ref{sec3}, we discuss how the theory is promoted to a gauge theory by the introduction of dynamical gauge fields. Moreover, we reformulate the gauge theory to match it with the degrees of freedom available in cold atomic gases and propose a possible implementation in a Bose-Fermi mixture in an optical lattice. We give an intuitive interpretation of various processes appearing in the proposed experimental setup, describe the envisioned experimental protocol, and discuss possible limitations. 
In Sec.~\ref{sec5}, we benchmark the proposed experiment with the example of the Schwinger mechanism. In particular, we show that an experiment with realistic parameters may extract the rate of particle--anti-particle production. 
Section~\ref{sec6} presents our conclusions. 
Three appendices give details on experimental implementation and parameters as well as numerical benchmark simulations.

\section{Wilson fermions}\label{sec2}

Before turning to dynamical gauge theories describing the Lorentz-invariant interaction of fermionic matter with gauge bosons,  we momentarily drop the gauge degrees of freedom for clarity. The non-interacting fermion part of the theory is described in the continuum in $d$ spatial dimensions by the  Dirac Hamiltonian
\begin{align}\label{eq:Dirac_Hamiltonian}
H_D = \int d^d \mathbf{x} \, \psi^\dagger (\mathbf{x}) \gamma^0 \left[i \gamma^j \partial_{j} + m\right] \psi (\mathbf{x}) \; ,
\end{align} 
where $\psi(\mathbf{x})$ is a fermionic Dirac spinor with $2^{d/2}$  components for $d$ even and $2^{(d+1)/2}$  components for $d$ odd. 
The $\gamma^0$ and $\gamma^j$ denote the gamma matrices in $d+1$ space-time dimensions~\cite{brauer1935spinors} and $\partial_j$ is a partial derivative in the spatial direction $j = 1, \dots, d$.
This Hamiltonian describes the kinetic energy and rest mass $m$ of Dirac fermions and leads to the dispersion relation of relativistic particles with energy $ \sqrt{m^2+\mathbf{p}^2}$. 

\subsection{The doubling problem}

For simulations on classical computers as well as on quantum devices consisting of sites in optical lattices or arrays of qubits, the continuum theory has to be discretized on a lattice. 
In view of quantum simulation, we work here in the Hamiltonian lattice formalism, with spatial lattice spacing $a$ and continuous {\em real} time $t$.

\begin{figure}
	\centering{\includegraphics[scale=0.4]{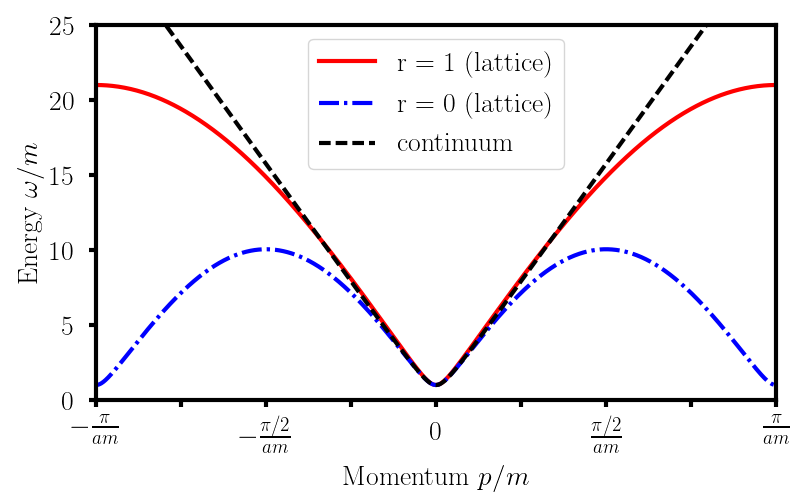}
		\caption{\label{fig:dispersion} 
			Comparison of the continuum and lattice dispersion relations, plotted within the first lattice Brillouin zone for lattice spacing $a = 0.1/m$ and $d=1$ spatial dimension. 
			The black, dashed line is the continuum result given by the Dirac Hamiltonian \eqref{eq:Dirac_Hamiltonian}, the low-energy behavior of which we aim at reproducing. 
			Discretizing the continuum kinetic energy by a nearest-neighbor hopping yields the blue, dashed-dotted dispersion relation, which has spurious low-energy states. 
			The addition of a Wilson term, see Eq.~\eqref{wilsonH}, removes the minima at the edge of the Brillouin zone, and thus effectively eliminates the fermion doublers (red, solid line; plotted here for $r = 1$). 
			}}
\end{figure}

The simplest discretization of fermions replaces the kinetic energy in Eq.~\eqref{eq:Dirac_Hamiltonian} with a nearest-neighbor hopping term. 
This naive procedure, however, leads to a discretized model with an additional ``doubling symmetry''~\cite{karsten1981lattice}. Its physical consequence is the appearance of spurious states, where each fermion in the continuum theory leads to $2^d$ fermion species for $d$ discretized dimensions, see Fig.~\ref{fig:dispersion}. 
These additional degrees of freedom affect the extrapolation to the continuum limit such that the correct continuum results are not recovered. 

The Nielsen--Ninomiya theorem \cite{nielsen1981absence} implies that, in order to remove these doublers, one has to sacrifice at least one of several fundamental characteristics of the continuum derivative: hermiticity, locality, discrete translational symmetry or chiral symmetry.
The choice, which of these characteristics to sacrifice, gives room for various strategies, with the staggered fermion~\cite{susskind1977lattice} and Wilson fermion~\cite{wilson1974confinement} prescriptions being among the ones most commonly known.

\subsection{Non-interacting Wilson fermions}

Wilson fermions sacrifice chiral symmetry to decouple the doublers from the low-energy degrees of freedom that describe the continuum theory. 
Despite the lack of chiral symmetry, relevant real-time phenomena can be efficiently simulated with Wilson fermions, with accuracy comparable to alternative implementations that respect chiral symmetry \cite{mace2017nonequilibrium}.

Wilson fermions can be understood  as the addition of a second-order derivative, which is discretized as 
\begin{align}
-\frac{a}{2} &\int d^{d}\mathbf{x} \sum_j  \, \psi^\dagger (\mathbf{x}) \gamma^0 \partial_j^2 \psi (\mathbf{x}) \\
&\rightarrow -\frac{a}{2} \sum_{\mathbf{n}}a^d \left(\psi_1\mathbf{n}^\dagger \gamma^0 \sum_j \frac{\psi_{\mathbf{n}+\mathbf{e}_j}  -2\psi_{\mathbf{n}} + \psi_{\mathbf{n}+\mathbf{e}_j} }{a^2} \right) \; . \nonumber
\end{align}
Here and in the following, the fermionic Dirac spinor $\psi_\mathbf{n}$ 
is located at lattice site $\mathbf{n} = \left(n_j\right)_{j=1,\dots,d}$ and $\mathbf{e}_j$ denotes a translation by a single site along the spatial direction $j$. 
This so-called Wilson term together with a hermitian discretization of the Dirac Hamiltonian \eqref{eq:Dirac_Hamiltonian} yields the lattice Hamiltonian of Wilson fermions,
\begin{align}\label{wilsonH}
H_W &= \sum_{\mathbf{n}} a^d \psi_\mathbf{n}^\dagger \gamma^0\left(m+ \frac{d\, r}{a}\right)\psi_\mathbf{n} \\
&\quad - \frac{a^{d-1}}{2}\sum_{\mathbf{n}}\left(\psi_\mathbf{n}^\dagger \gamma^0 \sum_j\left[i\gamma^j + r \right]\psi_{\mathbf{n}+\mathbf{e}_j} + \text{h.c.}\right) \; .\nonumber
\end{align}
The Wilson term is proportional to the lattice spacing $a$ and thus does not contribute in the continuum limit $a \rightarrow 0$.
Nevertheless, for any $a > 0$ it suppresses the fermion doublers as low-energy degrees of freedom, which is illustrated in Fig.~\ref{fig:dispersion}.  
This assures the recovery of the relevant continuum theory in the limit $a \rightarrow 0$. 
The strength of the Wilson term, given by the Wilson parameter $r$, can be adjusted in the range $0 < |r| \le 1$ while still describing the same continuum theory~\cite{montvay1997quantum}. In the following, we will exploit this freedom to optimize the implementation in a cold-atom quantum simulator.

\subsection{Optimized Hamiltonian}
In principle, the matrices $i \gamma^0\gamma^j + r\gamma^0$ in Eq.\ \eqref{wilsonH} couple all components of the spinors on neighbouring lattice sites. 
In an optical-lattice implementation, each of these couplings needs to be realized by a separate hopping process, which in the full gauge theory discussed below will moreover require the correct interactions with the gauge fields. 
With regard to experimental feasibility, it is thus highly desirable to minimize the number of coupling terms. In view of our application to Schwinger pair production below, we consider the case of one spatial dimension, $d=1$, where the gamma matrices can be represented with the three Pauli matrices $\sigma^\alpha$, $\alpha=x,y,z$,
\begin{align}
\gamma^0 = \sigma^\alpha \; , && \gamma^1 = i \sigma^\beta \; , && (\alpha \neq \beta) \; .
\end{align}
In the following, we choose $\alpha = x$ and $\beta = z$. In the Hamiltonian \eqref{wilsonH}, the second term then takes a particularly simple form by adjusting the Wilson parameter to $r=1$, such that
\begin{align}
\label{eq:gammas}
\gamma^0 = \begin{pmatrix}
0 & 1 \\ 1 & 0
\end{pmatrix} \; , && \gamma^1 = \begin{pmatrix}
i & 0 \\ 0 & -i
\end{pmatrix} \; , && i\gamma^0 \gamma^1 +r \gamma^0 = \begin{pmatrix}
0 & 2 \\ 0 & 0
\end{pmatrix}
\end{align}
These choices lead to the Hamiltonian
\begin{align}\label{1d_Wilson_Hamiltonian}
H_W &= \left(m + \frac{1}{a}\right) \sum_n \left(\psi_{n,1}^\dagger \psi_{n,2} + \text{h.c.}\right) \nonumber\\
&\quad + \frac{1}{a} \sum_n \left(\psi_{n,1}^\dagger \psi_{n+1,2} + \text{h.c.}\right) \; ,  
\end{align}
where we have substituted $\psi_n \rightarrow  \sqrt{a} \left(-1\right)^n \psi_n$ and written out the components, $\psi_n = \left(\psi_{n,1}, \psi_{n,2}\right)$, which fulfill the anti-commutation relations \begin{align}
\left\lbrace\psi_{n,\alpha}, \psi_{n',\beta}^\dagger \right\rbrace = \delta_{\alpha \beta} \delta_{nn'} \; .
\end{align}
The above choice of the $\gamma$-matrix representation and Wilson parameter is optimal in the sense that only one out of four possible terms that couple neighbouring lattice sites remains. 

Though the striking simplicity of \eqref{1d_Wilson_Hamiltonian} is special for one spatial dimension, in higher dimensions one can apply the same strategy of choosing different values of $r$, representations of $\gamma^\mu$, and canonical transformations to $\psi_\mathbf{n} \rightarrow C_{\mathbf{n}}\psi_{\mathbf{n}}$ to optimize for experimental needs. 

\subsection{Comparison to staggered fermions}

It is instructive to compare the above Wilson Hamiltonian to the lattice Hamiltonian of staggered fermions as it has been used in previous proposals for quantum simulators of lattice gauge theories. 
Staggered fermions sacrifice the discrete translational invariance on the lattice. They give a particularly simple formulation in one spatial dimension, where they enable one to analytically remove the fermion doublers~\cite{banks1976strong}. 
The Hamiltonian for non-interacting fermions reads in this case
\begin{align}\label{staggered_Hamiltonian}
H_{\text{st}} = \sum_n \left\lbrace m\left(-1\right)^n  c_n^\dagger c_n - \frac{i}{2a} \left[c_n^\dagger c_{n+1} - \text{h.c.}\right]\right\rbrace \; .
\end{align}
Here, the Dirac spinor is decomposed onto neighboring lattice sites such that there is only one fermionic degree of freedom $c_n$ living on each site.

In contrast, the Wilson formulation contains two components $\psi_{n,1}$ and $\psi_{n,2}$ at each lattice site, and thus realizes the same number of degrees of freedom in only half the space. Moreover, as discussed in the next section, gauge fields enter only on links connecting different lattice sites, such that Wilson fermions require only about half the gauge degrees of freedom. 

\begin{figure}
\includegraphics[scale=0.22]{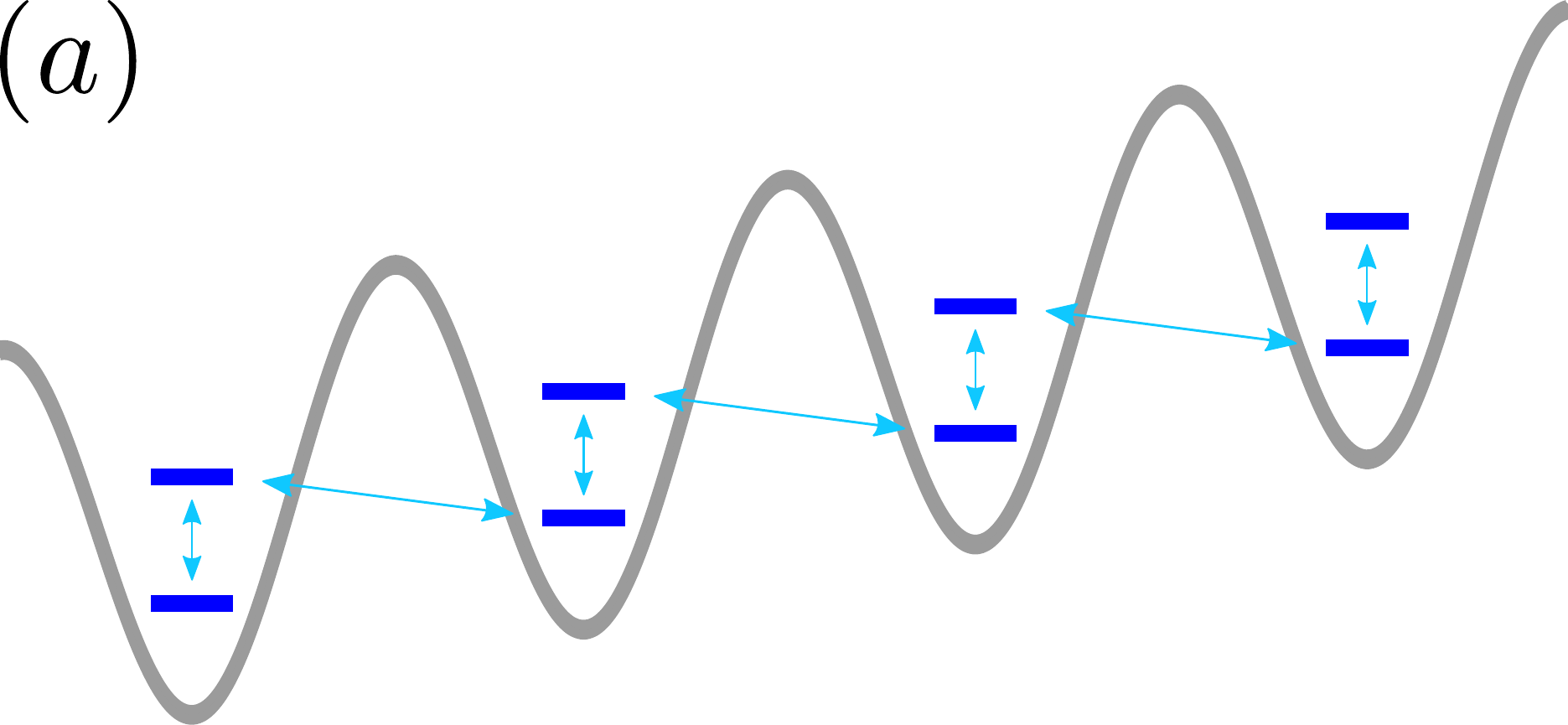}
\includegraphics[scale=0.22]{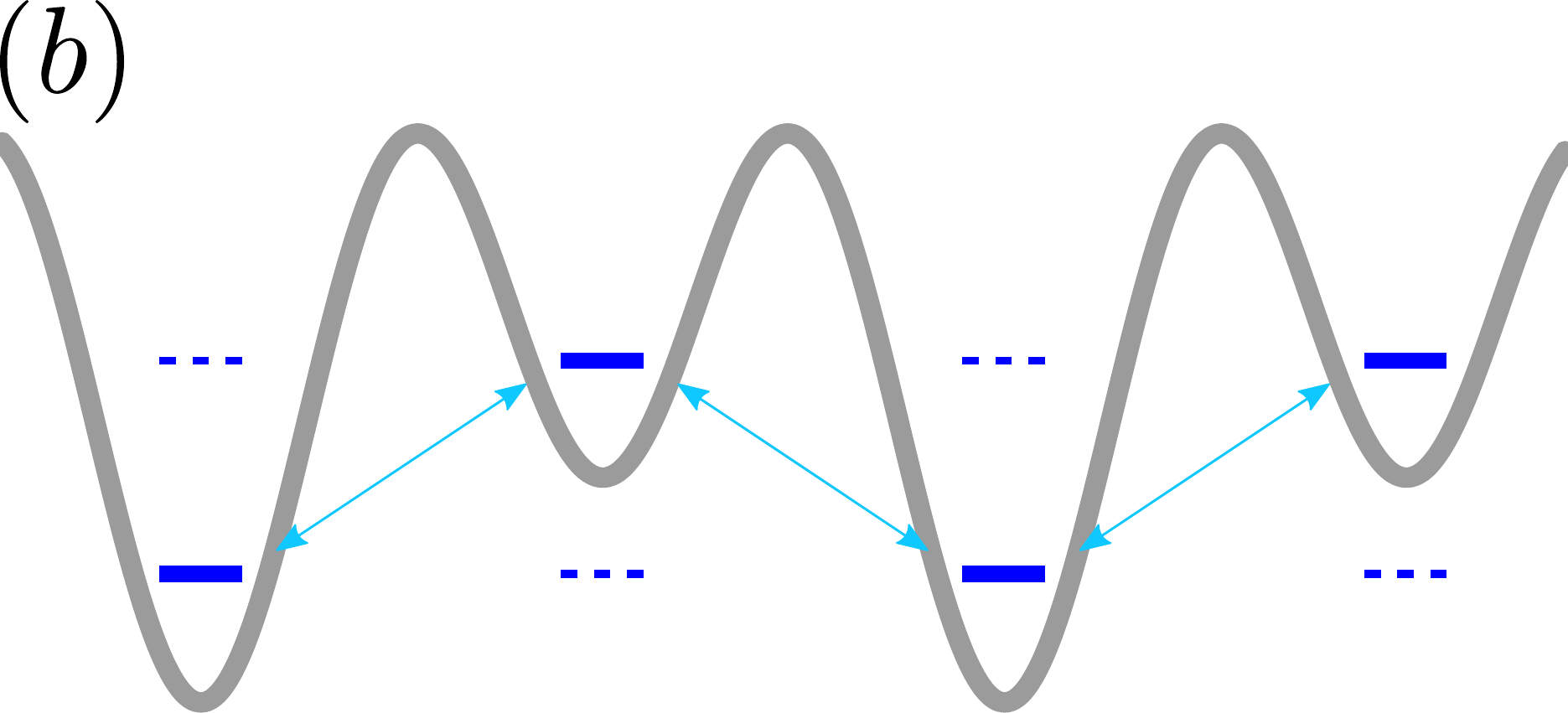}
\caption{\label{Wilson-vs-staggered} Sketch of a cold-atom implementation of lattice fermions in one spatial dimension. (a) Optimized Wilson formulation; (b) staggered formulation. The levels (blue) representing fermionic atoms are coupled as indicated with arrows. The grey curves represent optical potentials. Wilson fermions can be implemented in a tilted lattice, while proposals with staggered fermions typically require an optical superlattice, e.g. \cite{banerjee2012atomic}. 
}
\end{figure}

A prominent difference of Eq.~(\ref{staggered_Hamiltonian}) with respect to the optimized Wilson formulation (\ref{1d_Wilson_Hamiltonian}) concerns the sign factors appearing in the staggered mass term, which are also present in the interacting Hamiltonian that takes gauge fields into account. The experimental realization of the alternating on-site energy  typically requires an optical superlattice \cite{banerjee2012atomic}, while the Wilson formulation suggests a tilted potential as illustrated in Fig.~\ref{Wilson-vs-staggered}. The tilted potential could be less demanding experimentally and can be used to suppress unwanted tunneling processes as discussed in detail in Section~\ref{sec3}. 

Moreover, the decoupling of fermion doublers in the staggered formulation is special to $1+1$ space-time dimensions \cite{montvay1997quantum}. 
The theory becomes considerably more involved in higher dimensions where multiple, coupled fermion species have to be simulated. 
The resulting theory involves several coupled fermion species called ``tastes'', and the computation of physical observables requires a correction scheme by taking roots of the staggered fermion determinant. This ``rooting procedure'' is sometimes discussed controversially, but in practice many complications come from the multi-parameter fitting procedures that are required because of (``taste'') symmetry violations involving the spurious degrees of freedom in staggered formulations~\cite{montvay1997quantum}. 

In comparison, the Wilson decoupling of spurious doublers proceeds along the same lines in one or more spatial dimensions. 
On the other hand, while the low-energy sector of staggered fermions produces the correct dispersion relation up to order $a^2$, the usual Wilson Hamiltonian is only accurate to first order in the lattice spacing $a$. Nevertheless, as it has been recently shown, relatively simple (``tree-level'') improvements enable a remarkably good scaling towards the continuum limit of relevant real-time processes of QED and QCD in three spatial dimensions~\cite{mueller2016anomaly,muller2016chiral, mace2017simulating}. These include not only Schwinger pair production but also, e.g.\ the important phenomenon of anomalous currents due to the presence of quantum anomalies in QED and QCD. The latter depend crucially on the chiral characteristics of the system, showing that the explicit breaking of chiral symmetry by Wilson fermions is no fundamental roadblock.

\subsection{Coupling to gauge fields}

The strong potential of Wilson fermions for atomic quantum simulations becomes fully apparent when considering interacting gauge theories, as we will discuss now. Prominent examples for interacting gauge theories include QED based on the Abelian $U(1)$ gauge group, and QCD with underlying non-Abelian $SU(3)$ gauge group.~\cite{montvay1997quantum}. 

To illustrate the use of Wilson fermions, we proceed by focusing on the relatively simple example of the gauge group $\mathcal{G}=U(1)$ as realized in QED, where the fermionic matter fields $\psi_\mathbf{n}$ represent single-flavor Dirac spinors. In this case, since the $U(1)$ gauge group is local, gauge transformations amount to multiplication of the fermion fields with a potentially site-dependent phase $\alpha_\mathbf{n}$,
\begin{align} \label{gauge_trf}
\psi_\mathbf{n} &\rightarrow  e^{i\alpha_\mathbf{n}} \psi_\mathbf{n} \; .
\end{align}
Hopping terms such as $\psi_{\mathbf{n}}^\dagger \psi_{\mathbf{n}+\mathbf{e}_j}$ appearing in the non-interacting fermion Hamiltonian \eqref{wilsonH} are not invariant under this local gauge transformation for arbitrary non-constant $\alpha_\mathbf{n} \neq \alpha_{\mathbf{n}+\mathbf{e}_j}$.

In the interacting gauge theory, gauge invariance is obtained by coupling to operators $U_{\mathbf{n},j}$, which reside on the links between two neighbouring lattice sites $\mathbf{n}$ and $\mathbf{n}+\mathbf{e}_j$, and which transform as 
\begin{align}\label{gauge_trf_2}
U_{\mathbf{n}, j} \rightarrow e^{i\alpha_\mathbf{n}}U_{\mathbf{n}, j}e^{-i\alpha_{\mathbf{n}+\mathbf{e}_j}} \; .
\end{align}
The transformations \eqref{gauge_trf} and \eqref{gauge_trf_2} can be realized by the unitary operator $V = \exp\left[{ia^{d+1}\sum_{\mathbf{n}}\alpha_{\mathbf{n}}G_\mathbf{n}}\right]$ with the hermitian generator $G_\mathbf{n} = \sum_j\left( E_{\mathbf{n},j} -  E_{\mathbf{n}-\mathbf{e}_j,j}\right) + e \psi_\mathbf{n}^\dagger \psi_\mathbf{n}$. Here, $E_{\mathbf{n},j}$ is the conjugate field to $U_{\mathbf{n},j}$, which fulfills the commutation relation 
\begin{align}
\label{eq:commutatorEU}
\left[E_{\mathbf{n},j}, U_{\mathbf{m},k}\right] = e \delta_{jk}\delta_{\mathbf{n}, \mathbf{m}}U_{\mathbf{m},k} \; .
\end{align}
Gauge invariance corresponds to $\left[G_{\mathbf{n}}, H\right] = 0$. Thus, the full Hilbert space can be decomposed into sectors corresponding to different eigenvalues $q_{\mathbf{n}}$ of $G_{\mathbf{n}}$, and the physical Hilbert space is defined by picking a suitable subspace. 
Physically, the $q_{\mathbf{n}}$ are interpreted as the conserved charges of the group $\mathcal{G} = U(1)$, i.e.\ the electric charge. 
The restriction of the accessible Hilbert space to a single eigensector of the generator $G_{\mathbf{n}}$ is the lattice analogue of the familiar Gauss' law, which states that the electric charge is locally conserved, i.e.\ $\nabla E = \rho$. 
To fulfill the requirement of gauge invariance, in the interacting theory the hopping terms in the lattice Hamiltonian \eqref{wilsonH} are replaced by the combination 
$\psi_{\mathbf{n}}^\dagger U_{\mathbf{n},j} \psi_{\mathbf{n}+\mathbf{e}_j}$, which couples the dynamics of the fermions to that of the gauge fields.

With the presence of the gauge fields is associated an energy cost, governed by the electric Hamiltonian
\begin{align}
H_E = \frac{a}{2} \sum_{\mathbf{n},j} E_{\mathbf{n},j}^2\,.
\end{align}
This Hamiltonian is gauge invariant and implements the equations of motion for $U$ that give rise to the correct continuum limit as $a \rightarrow 0$ \cite{kogut1975hamiltonian}. 
In spatial dimensions higher than $d=1$, the gauge fields also have a magnetic contribution $H_B$, for details on which we refer to Refs.~\cite{kogut1975hamiltonian}. 
In total, we end up with the lattice Hamilonian of QED with Wilson fermions as 
\begin{align}
\label{eq:H_QED}
H_\text{QED} &= H_E + H_B +\sum_{\mathbf{n}} a^d \psi_\mathbf{n}^\dagger \gamma^0\left(m+ \frac{r}{a}\right) \psi_\mathbf{n} \\
& - \frac{a^{d-1}}{2}\sum_{\mathbf{n}}\left(\psi_\mathbf{n}^\dagger \gamma^0 \sum_j\left[i\gamma^j + r \right]U_{\mathbf{n},j}\psi_{\mathbf{n}+\mathbf{e}_j} + \text{h.c.}\right). \nonumber
\end{align}

As far as the fermion sector is concerned, it is straightforward to generalize the above construction also to non-Abelian gauge theories. 
In the fundamental representation of $SU(N)$, the fermion spinors carry an additional group index and the link variables require a different formulation, but the structure of the gauge-matter interactions as given by Eqs.~\eqref{gauge_trf} and \eqref{gauge_trf_2} remains the same~\cite{montvay1997quantum}.
For staggered fermions, implementations of non-Abelian gauge theories with cold atoms have been discussed in Refs.~\cite{banerjee2013atomic,zohar2013cold,rico2018so}.

\section{Cold-atom QED \label{sec3}}

The lattice gauge theory written in Eq.~\eqref{eq:H_QED} consists of fermions interacting with gauge fields. We now reformulate the theory in a way that matches the degrees of freedom available in cold atomic gases, using the simplest case of QED in one spatial dimension, also known as the massive Schwinger model \cite{coleman1976more}. In this case, Eq.~\eqref{eq:H_QED} simplifies due to the absence of the magnetic field term $H_B$ and we can adopt the compact form of Eq.~\eqref{1d_Wilson_Hamiltonian}. 

\subsection{Optimized cold-atom Wilson Hamiltonian}

Using the same optimization choices as led to Eq.~\eqref{1d_Wilson_Hamiltonian}, Eq.~\eqref{eq:H_QED} 
yields the quantum many-body Hamiltonian
\begin{align}
\label{eq:H_QED_1D}
H_\text{QED} &=  \sum_n \left\lbrace \frac{a}{2} E_n^2 + \left(m+\frac{1}{a}\right) \psi_n^\dagger \begin{pmatrix}
0 & 1 \\ 1 & 0
\end{pmatrix}
\psi_n \right\rbrace \nonumber\\&+ \frac{1}{a} \sum_n \left\lbrace \psi_n^\dagger \begin{pmatrix}
0 & 1 \\ 0 & 0 
\end{pmatrix}
U_n\psi_{n+1} + \text{h.c.} \right\rbrace \; .
\end{align}
To alleviate notation, we label the gauge fields for $d=1$ only with the site to the left, i.e.\ $E_n=E_{n,j=1}$ and analogously for $U$.  Here, $n=1\dots N$ is the number of lattice sites. 
Thanks to the choices of the previous section, the number of links carrying the gauge--matter interactions is only half that of staggered fermions for the same number of quantum simulated fermionic degrees of freedom.

The generators of the gauge transformations are now given by
\begin{align}
G_n = E_n- E_{n-1} - e \sum_{\alpha=1,2} \psi_{n,\alpha}^\dagger  \psi_{n,\alpha} \; .
\end{align}
We choose our physical states from the zero-charge sector $G_n |\text{phys}\rangle = 0$. 

The remaining procedure is similar to previous implementations with staggered fermions \cite{kasper2016schwinger,kasper2017implementing}. 
The commutation relation \eqref{eq:commutatorEU}, together with the requirements of $E$ being hermitian and $U$ being unitary, can only be fulfilled in an infinite-dimensional Hilbert space. Since the quantum control of infinitely many degrees of freedom is in practice impossible, we adopt the so-called quantum link \cite{chandrasekharan1997quantum} formalism as a regularization, which replaces the gauge operators on each link by spin operators,
\begin{align}\label{spin_replacement}
E_n \rightarrow e L_{z,n} \; , && U_n \rightarrow \left[\ell (\ell+1)\right]^{-1/2} L_{+,n} \; ,
\end{align}
with $\left[L_{n,\alpha},L_{m,\beta}\right] = i \delta_{nm}\epsilon_{\alpha \beta \gamma}L_{m,\gamma} \;, \;   \alpha, \beta, \gamma \in \left\lbrace x, y, z \right\rbrace$ and $L_{\pm,n} = L_{x,n} \pm i L_{y,n}$.
This regularization leaves gauge invariance and the commutation relation \eqref{eq:commutatorEU} intact, but sacrifices unitarity of the link operators, which now fulfill the commutation relation $\left[U_n, U_m^\dagger\right] = 2\delta_{nm}E_m/\left[e\ell(\ell+1)\right]$. Already for small representations of the quantum spin, quantum link models share salient features with QED, such as confinement and string breaking \cite{banerjee2013atomic}. Moreover, in the limit of large spins $(\ell \rightarrow \infty)$, which we are focussing on, one recovers full QED.  
Finally, we represent the spin operators with two Schwinger bosons,
\begin{align} L_{z,n} = \frac{1}{2} \left(b_n^\dagger b_n - d_n^\dagger d_n\right) \; , &&
L_{+,n} = b_n^\dagger d_n \; ,
\end{align}
which fulfill the constraint
\begin{align}\label{Schwinger_boson_constraint}
2\ell = b_n^\dagger b_n + d_n^\dagger d_n \; .
\end{align}
This yields the final Hamiltonian that may be realized with cold atoms in an optical lattice,
\begin{align}\label{cold_atom_H}
H_\text{CA} &=  \frac{ae^2}{4} \sum_n   \left(  b_n^\dagger b_n^\dagger b_n b_n + d_n^\dagger d_n^\dagger d_n d_n\right) \nonumber\\ &\quad+ \left(m+\frac{1}{a}\right) \sum_n\left(\psi_{n,1}^\dagger 
\psi_{n,2} + \text{h.c.}\right)  \\ &\quad+ \frac{1}{a\sqrt{\ell (\ell+1)}}\sum_n \left( \psi_{n,1}^\dagger 
b_n^\dagger d_n \psi_{n+1,2} + \text{h.c.}\right) \; ,\nonumber
\end{align}
where we have dropped an irrelevant constant after using that $L_{z,n}^2 = \left(b_n^\dagger b_n - \ell\right)^2\! /2 + \left(\ell -d_n^\dagger d_n\right)^2\! /2$. As it becomes apparent in this formulation, gauge degrees of freedom only enter in couplings between matter fields at different sites, but not in the on-site terms $\sim\psi_{n,1}^\dagger \psi_{n,2}$.

\subsection{Experimental Implementation\label{sec4}}
\begin{figure}
\includegraphics[scale=0.6,angle=-90]{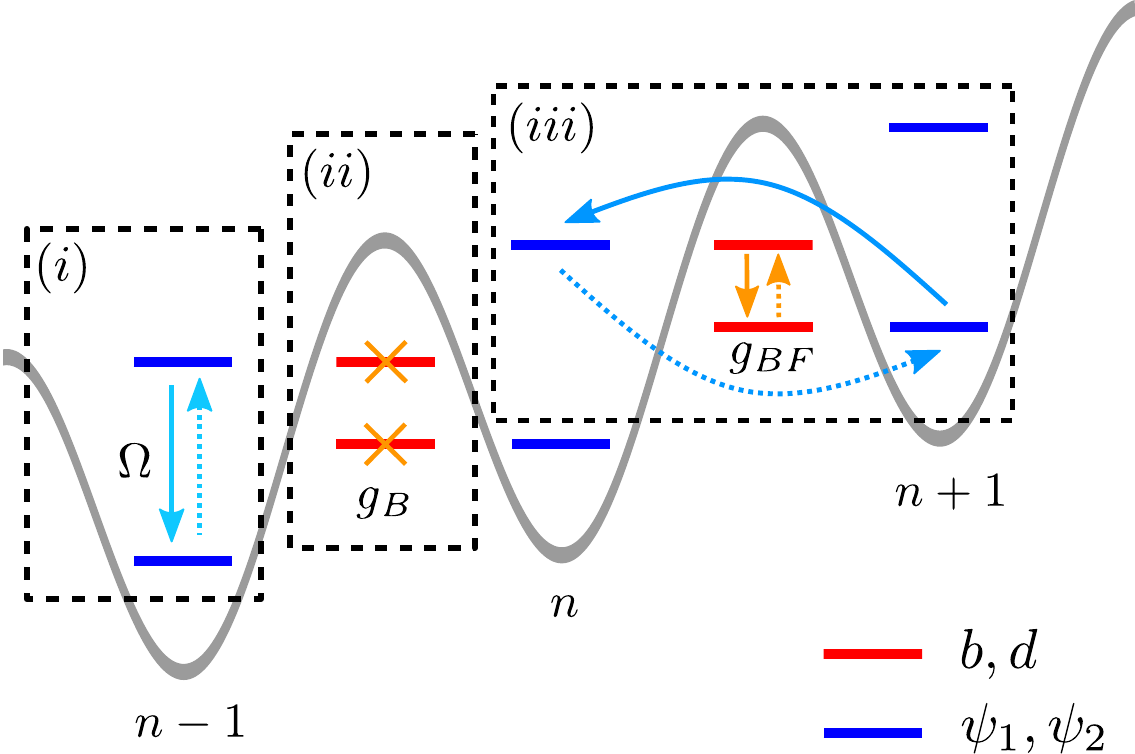}
\caption{\label{tilted_lattice}Sketch of the proposed implementation. 
Fermions and bosons (blue and red bars, respectively) can occupy two hyperfine states each and are trapped in a tilted optical lattice potential (grey line). 
The gauge-invariant dynamics of 1+1D QED is realized by three processes: 
$(i)$ local oscillation between the fermionic species [see Eq.~\eqref{omega}]; 
$(ii)$ local bosonic self-interaction [see Eq.~\eqref{g_boson}]; 
$(iii)$ correlated hopping of both species due to spin changing collisions [see Eq.~\eqref{g_SCC}].
}
\end{figure}

We propose to realize the Hamiltonian \eqref{cold_atom_H} with a Bose-Fermi mixture in a tilted optical lattice as sketched in Fig.~\ref{tilted_lattice}. Transverse motion is frozen out by a strong radial confinement, rendering the system effectively one-dimensional. The mixture is additionally subjected to an optical lattice potential that is attractive (repulsive) for the fermions (bosons), such that the atomic species are allocated in an alternating fashion. In the following, we will refer to the positions of the fermions  (bosons) as sites (links). 
For a sufficiently deep lattice, the atoms will occupy localized Wannier states, such that tunneling beyond neighboring sites (links) can be neglected. Tilting the optical potential suppresses this direct tunneling and effectively localizes the atoms on single sites (links). 
Moreover, the species are prepared in two selected hyperfine states each (denoted by annihilation operators $\psi_{1,n}, \psi_{2,n}$ respectively $b_n,d_n$). 
The desired dynamics governed by Eq.\ \eqref{cold_atom_H} can now be realized by implementing the following three interactions among these states (angular momentum and energy conservation ensure that the dynamics accesses no other states \cite{zohar2013quantum,kasper2017implementing}).

\begin{enumerate}
	\item[(i)] We propose to drive resonant oscillations with Rabi frequency $\Omega$ between the two fermionic states on each site using radiofrequency radiation. 
	This realizes the second line of Hamiltonian \eqref{cold_atom_H}, i.e.\ the electron mass plus the on-site part of the Wilson term, 
	\begin{align}\label{omega}
	\Omega \leftrightarrow m + \frac{1}{a} \; .
	\end{align}
	\item[(ii)] We assume $2\ell$ Bose condensed atoms on each link. In the trapping potential, the atomic cloud feels an effective interaction constant $g_B$ given by the scattering length, the boson mass, and overlap integrals over localized Wannier functions, see Eq.~\eqref{eq:gB}. These interactions set the energy scale for the simulated electric field, the first line of Hamiltonian \eqref{cold_atom_H}, 
	\begin{align}\label{g_boson}
	g_B  \leftrightarrow \frac{ae^2}{2} \; .
	\end{align}	
	\item[(iii)] The gauge-invariant interactions in the third line of Hamiltonian \eqref{cold_atom_H} are implemented using spin changing collisions (SCC) between the fermions and bosons \cite{zohar2013quantum,kasper2017implementing,stannigel2014constrained}. For this purpose, one has to choose appropriate hyperfine states to ensure angular momentum conservation and then apply an external magnetic field in order to tune the SCC into resonance. These lead to a correlated hopping of the bosons and fermions with an effective interaction $g_{BF}$, see Eq.~\eqref{eq:g_BF}. 
	These interactions set the lattice-spacing parameter of the quantum-simulated gauge theory, 
	\begin{align}\label{g_SCC}
	g_{BF}  \leftrightarrow \frac{1}{a \sqrt{\ell(\ell + 1)}} \; .
	\end{align}
\end{enumerate}
For further details on a possible realization, we refer to the appendix \ref{app:A}.
Equations (\ref{omega}-\ref{g_SCC}) define the two relevant dimensionless parameters of the simulated theory, $am$ and $e/m$. Note that the value of $a$ is not equivalent to the optical-lattice spacing $a_{\mathrm{lat}}$ imposed in the quantum simulator [see Eq. \eqref{optical_potential}].

\subsection{Interpretation of the cold atom Hamiltonian}
The individual processes contributing to Eq.~\eqref{cold_atom_H} permit of physical interpretations in simplified limits, which are useful to gain some intuition. 

\subsubsection{Free fermion Hamiltonian} 
The fermionic part of Hamiltonian \eqref{cold_atom_H} becomes particularly simple in the absence of interactions with the gauge fields. 
Referring to the single-particle states of $\psi_{n,1}$ and $\psi_{n,2}$ as $\left|\uparrow \right\rangle_n$ and $\left|\downarrow\right\rangle_n$, respectively, we start by considering the local, purely fermionic part in the second line of \eqref{cold_atom_H}, which dominates in the heavy-mass limit $m\to\infty$. It is diagonal in the basis $\left|\leftarrow\right\rangle_n = \frac{1}{\sqrt{2}} \left(\left|\uparrow\right\rangle_n - \left|\downarrow\right\rangle_n\right)$ and $\left|\rightarrow\right\rangle_n = \frac{1}{\sqrt{2}} \left(\left|\uparrow\right\rangle_n + \left|\downarrow\right\rangle_n\right)$, with eigenvalues $-m$ and $+m$, respectively. In this basis the local fermionic Hilbert space is given by
\begin{align}
\mathcal{H}_n = \left\lbrace |0_\leftarrow0_\rightarrow \rangle_n, |1_\leftarrow 0_\rightarrow \rangle_n, | 0_\leftarrow 1_\rightarrow\rangle_n , |1_\leftarrow 1_\rightarrow \rangle_n \right\rbrace\; ,
\end{align}
where $|j_\rightarrow k_\leftarrow\rangle_n$ denotes a state with $j$ fermions in the state $\left|\rightarrow\right\rangle_n$  and $k$ fermions in $\left|\leftarrow\right\rangle_n$. We can therefore identify the fermionic vacuum and electron/positron states according to
\begin{subequations}\label{identification}
\begin{align}
\text{vacuum (``Dirac sea")}: &&|\Omega\rangle_n &\leftrightarrow | 1_\leftarrow 0_\rightarrow\rangle_n  \; ,\\
\text{electron}: &&|e^-\rangle_n &\leftrightarrow | 1_\leftarrow 1_\rightarrow\rangle_n \; ,\\
\text{positron}: &&|e^+\rangle_n &\leftrightarrow | 0_\leftarrow 0_\rightarrow\rangle_n \; ,\\
\text{electron + positron}: && |e^- e^+\rangle_n &\leftrightarrow |0_\leftarrow 1_\rightarrow \rangle_n \; .
\end{align}
\end{subequations}
Intuitively, an electron corresponds to the presence of a fermion in $\left|\rightarrow\right\rangle_n$, while a positron corresponds to the absence of a fermion in $\left|\leftarrow\right\rangle_n$.

For a finite fermion mass $m$, we have to take into account the fermionic hopping terms in \eqref{cold_atom_H}. The decomposition of the vacuum state into the local fermionic states is then formally given by a Slater determinant involving all lattice sites. In this case, it is more convenient to describe the quantum system in terms of correlation functions. In absence of interactions with the gauge field, the matter fields form a free theory. An initial vacuum of non-interacting fermions can thus be completely described in terms of the equal-time statistical propagators $F_{mn}^{\alpha \beta} =\frac{1}{2} \left\langle \left[ \psi_{m,\alpha} , \psi_{n,\beta}^\dagger \right] \right\rangle$. In momentum space, this non-interacting vacuum is characterized by the correlations (see also appendix \ref{app:C})
\begin{align}\label{vac_corrs}
F_{kk}^{11} = 0 \; , && F_{kk}^{22} = 0 \; , && F_{kk}^{21} = \frac{\omega_k}{2z_k}
\end{align}
and $F_{kk'}^{\alpha\beta}=0$ for $k\neq k'$. Here, $z_k  = m+\frac{1}{a} \left(1+ \exp \left(\frac{2\pi i k}{N}\right) \right)$ and the dispersion $\omega_k = |z_k|$, $k = 0,1, \dots , N-1$. 

\subsubsection{Gauge-field energy} 
In the experiment, the gauge part corresponds to an array of trapped spinor BECs in two hyperfine states. For the semi-classical limit of large occupation numbers, every local BEC can be pictured by a collective spin Bloch sphere. Since the simulated electric field corresponds to an occupation imbalance between the two states, it can be associated with the azimuthal angle measuring the distance from the equator of the Bloch sphere. Thus, the electric energy, which arises from the bosonic self-interactions in Eq.~\eqref{cold_atom_H} corresponds to the so-called one-axis twisting Hamiltonian \cite{kitagawa1993squeezed}. This clarifies the contribution from \eqref{g_boson}: It generates a rotation of the polar angle, whose frequency depends on the azimuthal angle. This corresponds to a phase rotation of $U$ in the gauge theory. However, this simple dynamics that happens locally on every link is modified by the correlated hopping of bosons and fermions. 

\subsubsection{Correlated hopping} 
In the heavy-mass limit underlying the identifications of equation \eqref{identification}, one can also easily visualize the effect of the correlated hopping \eqref{g_SCC}. For example, the elementary process for the local production of a single $e^+ e^-$ pair is composed of the hopping and simultaneous flipping of a single (fermionic) spin from one site to the next, while decreasing the (bosonic) imbalance on the link joining the two sites. 

As for the free-fermion part discussed above, the simplified interpretation in terms of local single-particle states is convenient to gain a basic understanding of the cold-atom system, but in order to describe the full complexity of the many-particle quantum dynamics it becomes necessary to consider many-body correlation functions. 
Indeed, for a finite fermion mass, pair production happens non-locally and can only be detected by measuring correlation functions. 
In fact, even the concept of a particle is ill-defined in the generic interacting non-equilibrium situation \cite{gelis2016schwinger}. As a measure for the total fermion particle number density $n = \frac{1}{L}\sum_k \tilde{n}_k$, we employ a typical definition following from the instantaneous diagonalization of the purely fermionic contribution to the full Hamiltonian. Then, $n$ can be expressed in terms of the energy density $\tilde{\epsilon}_k$, which is a function of the statistical propagator, and the dispersion $\tilde{\omega}_k$,
\begin{align}\label{number-corrs}
\tilde{n}_k = \frac{\tilde{\epsilon}_k}{\tilde{\omega}_k} + 1 \; , &&
\tilde{\epsilon}_k = -  \left( \tilde{z}_k F_{kk}^{21} + \left[\tilde{z}_k F_{kk}^{21}\right]^* \right) \; ,
\end{align}
where the tilde indicates that all quantities have to be calculated on the background of the gauge fields.

\subsection{Experimental limitations}
There are a number of experimental limitations that set bounds on the implementation of \eqref{cold_atom_H}. The most important ones limit the accessible time-scales in the experiment as follows [for further details, see the appendices \ref{app:A} and \ref{app:B}].

First of all, we have replaced the gauge fields by finite spin operators \eqref{spin_replacement}. To quantitatively approach QED predictions, we would thus like to employ BECs with large atom numbers corresponding to $\ell \rightarrow \infty$. 
The large boson density will lead to considerable three-body losses \cite{pethick2002bose}, which depend on the precise lattice structure. 
These losses set a limiting time $T_3$ for the validity of the quantum simulation. 

A second restriction comes from the need to suppress direct hopping terms of the two species and to conserve the boson number locally on each link to ensure the constraint \eqref{Schwinger_boson_constraint}. Wilson fermions naturally favour a tilted lattice, which conveniently suppresses direct tunneling. On the downside, the tilt renders states localized on single lattice sites unstable \cite{gluck2002wannier}. After a time $T_{LZ}$, they decay due to Landau-Zener transitions, which is the second main limitation of our proposed setup. 

Finally, experiments will implement a set of lattice QED parameters $a,m,e$ with a resulting Brillouin zone of finite size $\sim {1}/{a}$. Since we are interested in Schwinger pair production in strong electric fields, the creation and subsequent acceleration of particles becomes unphysical when the energy of these particles reaches the cutoff $\sim {1}/{a}$. This gives a third time-scale $T_{\mathrm{lat}}$ that limits the accessible dynamics.

An experimental implementation will have to carefully balance between these different imperfections. Nevertheless, as we will show in the remainder of this paper, the observation of relevant phenomena is achievable in state-of-the art experiments.
Moreover, $T_3$ and $T_{\mathrm{lat}}$ can be mitigated if we do not require quantitative agreement with continuum QED. For any finite $a$, the experiment will implement a valid lattice gauge theory. Similarly, by settling for finite representations $l<\infty$, one implements quantum links models, which are valid gauge theories in their own right. Already for extremely small representations, these share the most salient qualitative features with usual QED, such as string-breaking dynamics \cite{banerjee2012atomic}.

\section{Benchmark: Onset of Schwinger Pair Production\label{sec5}}
Having shown how to use Wilson fermions in order to realize a particularly compact formulation of 1+1D QED in optical lattices, we now turn to making quantitative predictions for the proposed experiment.
To this end, we perform numerical simulations of an important effect that occurs in 1+1D QED: pair production via the Schwinger mechanism \cite{sauter1931verhalten,schwinger1951gauge}. 
The mechanism describes how a strong external electric field transfers energy to vacuum fluctuations and turns them into pairs of real particles and anti-particles. 
In the present case of 1+1 dimensions, one can analytically compute the particle-production rate in the continuum
\begin{align}
\label{eq:SchwingerRate}
\frac{\dot{n}}{m^2} = \frac{E}{\pi E_c} \exp \left(- \frac{\pi E_c}{E}\right) \; ,
\end{align}
which is valid for a constant background field that was turned on in the infinite past \cite{schwinger1951gauge}.
Most importantly, the exponential factor induces a dramatic increase of particle production above a critical field $E_c = m^2/e$. 
In this regime, the particle production is non-perturbative and thus constitutes an excellent non-trivial target for quantum simulation. 
Moreover, there is no known analytic prediction for the fully interacting theory including the back-reaction of the produced particles onto the gauge fields \cite{gelis2016schwinger}.

In the following, we show that the proposed implementation strategy, using a mixture of fermionic ${}^6 \text{Li}$ and bosonic ${}^{23} \text{Na}$ based on current technology, allows one to simulate the Schwinger mechanism and the non-perturbative particle-production rate. For benchmarking purposes, we consider the limit of weak gauge coupling, where powerful functional-integral methods provide quantitatively reliable results \cite{kasper2014fermion,buyens2017real} (see appendix \ref{app:C} for a summary of the simulation procedure). Beyond this weak-coupling benchmark regime, the experiment should be able to proceed also into the regime of strong coupling, where these functional-integral methods are expected to fail.  

\subsection{Proposed experimental protocol}

To quantum simulate the Schwinger mechanism, we propose the following
experimental protocol. First, the bosonic and fermionic atoms are loaded into the tilted optical
lattice structure. At sufficiently low temperatures, we will have single
fermions per lattice site in the state $\left|\downarrow \right\rangle_n$
and single-component condensates with average particle number $\langle
d_n^\dagger d_n \rangle$ on the links. 
Initially, the imbalance $\delta N_n = \langle b_n^\dagger b_n - d_n^\dagger
d_n \rangle$ of the two hyperfine states, which defines the electric field,
is tuned to $\delta N_n=0$. 
This can be achieved with a linear coupling between the two states, e.g.\
via radiofrequency radiation. 
This initial state, where the electric field is prepared in a product of
coherent states, is only approximately restricted to a single gauge sector.
However, as it has been shown numerically \cite{kasper2014fermion}, in the
limit of large boson number and for the present scenario the small
fluctuations of the electric field are too insignificant to compromise the
gauge-invariant dynamics.

As described in the previous section, in the heavy-mass limit, the fermionic
vacuum is given by the local superpositions $\left|\leftarrow\right\rangle_n
= \frac{1}{\sqrt{2}} \left(\left|\uparrow\right\rangle_n -
\left|\downarrow\right\rangle_n\right)$. This state can be easily generated
from a polarized gas by a $\pi/2$ radiofrequency pulse (with a phase shift
of $\pi/2$ with respect to the radiofrequency pulse $\propto\Omega$ that
drives the dynamics). Together with the gauge fields, this realizes the
ground state at infinite fermion mass. Alternatively, one may adiabatically
prepare the ground state at a finite value of the fermion rest mass, e.g.\
by adiabatically ramping down the optical lattice such that the correlated
hopping is gradually turned on.

With either choice of the fermion initial state, the dynamics can be started
by quenching the bosonic imbalance $\delta N_n$ from zero to a desired
initial electric field. For the experimentally less demanding case of the
infinite-mass ground state, this will realize a combination of the targeted
Schwinger mechanism with a mass quench from infinity to a desired, finite
value. Below, we demonstrate that both initialization procedures yield
comparable results for the particle production rate, at least for the
parameter regimes studied in this work (i.e.\ weak coupling and relatively
coarse lattices).

After the initialization procedure, the system evolves for a desired time
under Hamiltonian \eqref{cold_atom_H}, after which we extract the relevant
observables. In particular, we are interested in the volume-averaged
electric field and the total fermion particle number. The first is easily
achieved by reading out the bosonic imbalance via standard absorption
imaging. Current experimental techniques allow for single-site resolution,
such that one can also access the local electric fields and their spatial
correlations.

The total fermion particle number can be measured in a number of ways. 
First, the definition of the particle number given in Eq.~\eqref{number-corrs}
can be measured by adiabatically transferring the system 
to the limit of infinite mass by increasing $\Omega$, see Eq.~\eqref{omega}. 
A subsequent $\pi/2$ radiofrequency pulse around the $y$ axis on the fermion 
Bloch sphere maps fermions in the upper (lower) band onto the pseudo-spin state $\left|\uparrow\right\rangle$ ($\left|\downarrow\right\rangle$).
Through a Stern-Gerlach measurement, one can thus detect the produced
particle--anti-particle pairs. In addition, their momentum dependence can be
resolved by time-of-flight imaging. 
Since the increase of $m$ amounts to an increase of the critical field
$E_c=m^2/e$, this scheme effectively turns the Schwinger mechanism off
smoothly, in a similar spirit as for the so-called Sauter pulses that are
often used to model the Schwinger mechanism in time-dependent (classical)
background fields \cite{gelis2016schwinger}. 
Second, one can map out the full band structure by adapting the tomography
scheme developed in Ref.~\cite{hauke2014tomography} and first demonstrated
experimentally in Ref.~\cite{flaeschner2016experimental}. 
The particle number is obtained by comparison to the tomography for the ground state at the mass parameter targeted in the dynamics. 
Third, the full information about the fermionic part of the theory can be reconstructed by measuring spin-dependent and spatially resolved fermion correlation functions, either via quantum-gas microscopy 
\cite{parsons2016site,boll2016spin,cheuk2016observation} or, since we are interested in the spatial continuum limit, on coarse grained length scales larger than the lattice spacing.

\subsection{Simulated real-time dynamics of fermion density and electric field}

\begin{figure}
\centering{\includegraphics[scale=0.35]{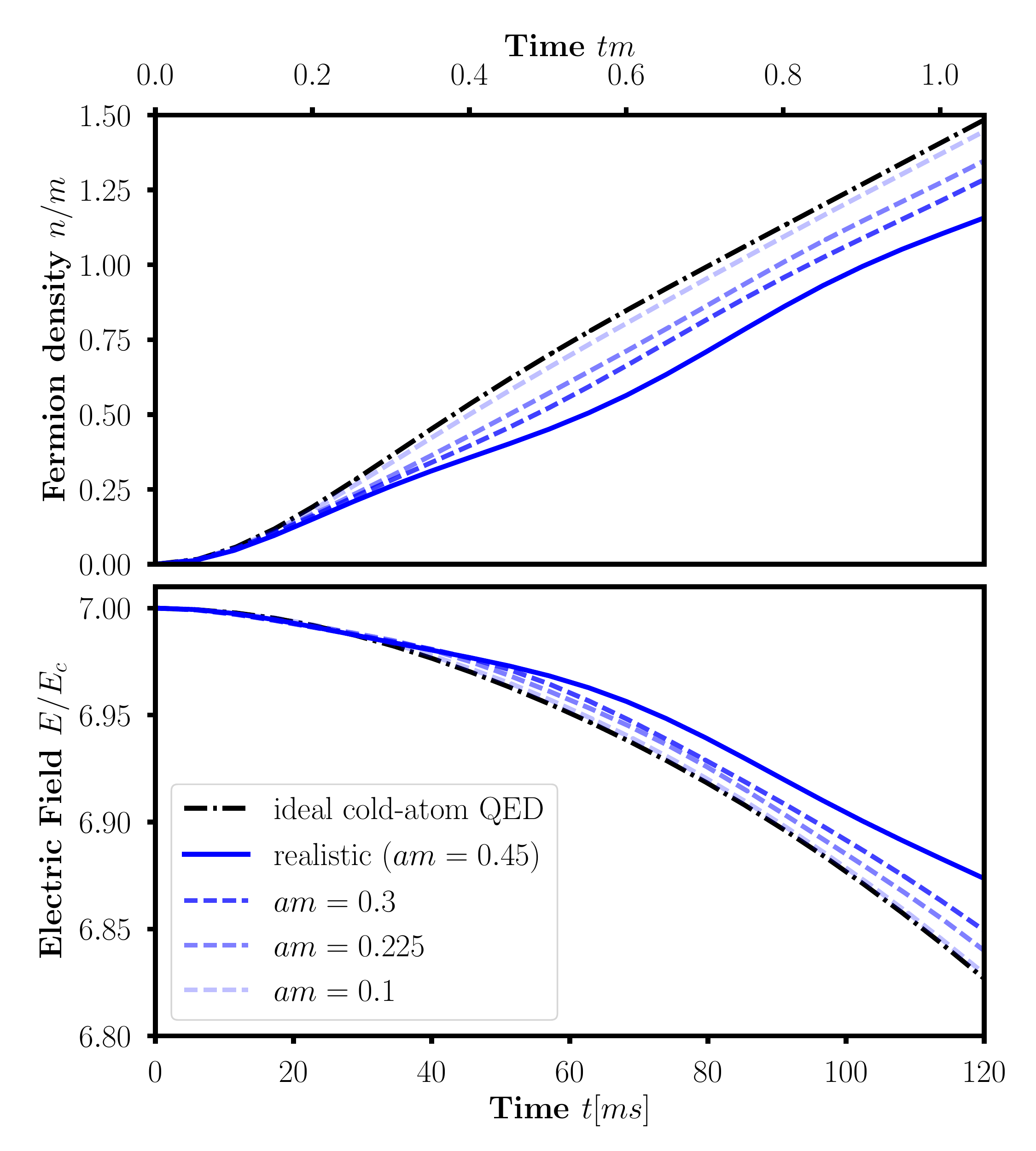}
\caption{\label{simulation2}
Benchmarking simulations ($e/m = 0.45$, $am=0.45$, $N=20$, and $\ell = 1500$; blue solid line) qualitatively recover the ideal continuum Schwinger rate (the limit of small $am$ as well as large $N$ and $\ell$; black dash-dotted). 
As the interpolating lines for different $am$ at fixed volume $L = aN = 9/m$ show, the benchmark simulations are converged with respect to $\ell$ and smoothly approach the continuum prediction (dashed lines from dark to light blue $am =0.3$, $0.225$, and $0.1$). 
}}
\end{figure}

While lattice-QED has only two free parameters $am$ and $e/m$, one can change various ingredients on the experimental side. Different choices can drastically affect the various time scales governing the quantum simulator. In appendix \ref{app:B}, we outline a simple optimization procedure. It leads us to two examples of realistic parameter sets, which translate to
\begin{align}
\text{(i):} && am &= 0.45 \;, && e/m = 0.45 \; , && E_0/E_c = 7  \; . \nonumber\\
\text{(ii):} && am &= 0.94 \;, && e/m = 0.22 \; , && E_0/E_c = 3  \; , \nonumber
\end{align}
As these values illustrate, if working at finer lattice spacing $am$, experimental restrictions of the present proposal require an increase of the coupling $e/m$. This combination limits the numerical technique employed here to coarse lattices, as it is quantitatively reliable only at small $e/m$.  
However, in the experiment it will allow one to reach exactly the most interesting regime, namely close to the continuum limit and with strong coupling.  

Since the results for these parameter sets are qualitatively very similar, we discuss in the following only the set (i) and delegate the results of set (ii) to the appendix (Fig.~\ref{simulation1}). For both choices, the initial electric field $E_0$ exceeds the critical value in Eq.~\eqref{eq:SchwingerRate}, and thus we expect an enhanced production of electron-positron pairs. 
We simulate the real-time evolution of the particle number as well as the decay of the initial electric field and compare them to an idealized implementation with $\ell,N\to\infty$ and $a\to 0$. 
For the presented realistic parameter sets, though quantitative deviations from the expected QED behaviour occur, the observed particle production shows good qualitative agreement. It is also possible to observe the onset of the decay of the electric field due to the backreaction of the fermions. 
Observed deviations are mainly due to the large lattice spacings and do not result from the finite $\ell < \infty$ (see interpolating curves with different $a$). The quantum simulator will thus be capable of simulating a dynamical lattice gauge theory, though the realistically reachable lattice parameters $a$ still lead to quantitative deviations from the spatial continuum limit, at least for the weak coupling regime.

\subsection{Non-perturbative particle production rate}

\begin{figure}
\centering{\includegraphics[scale=0.33]{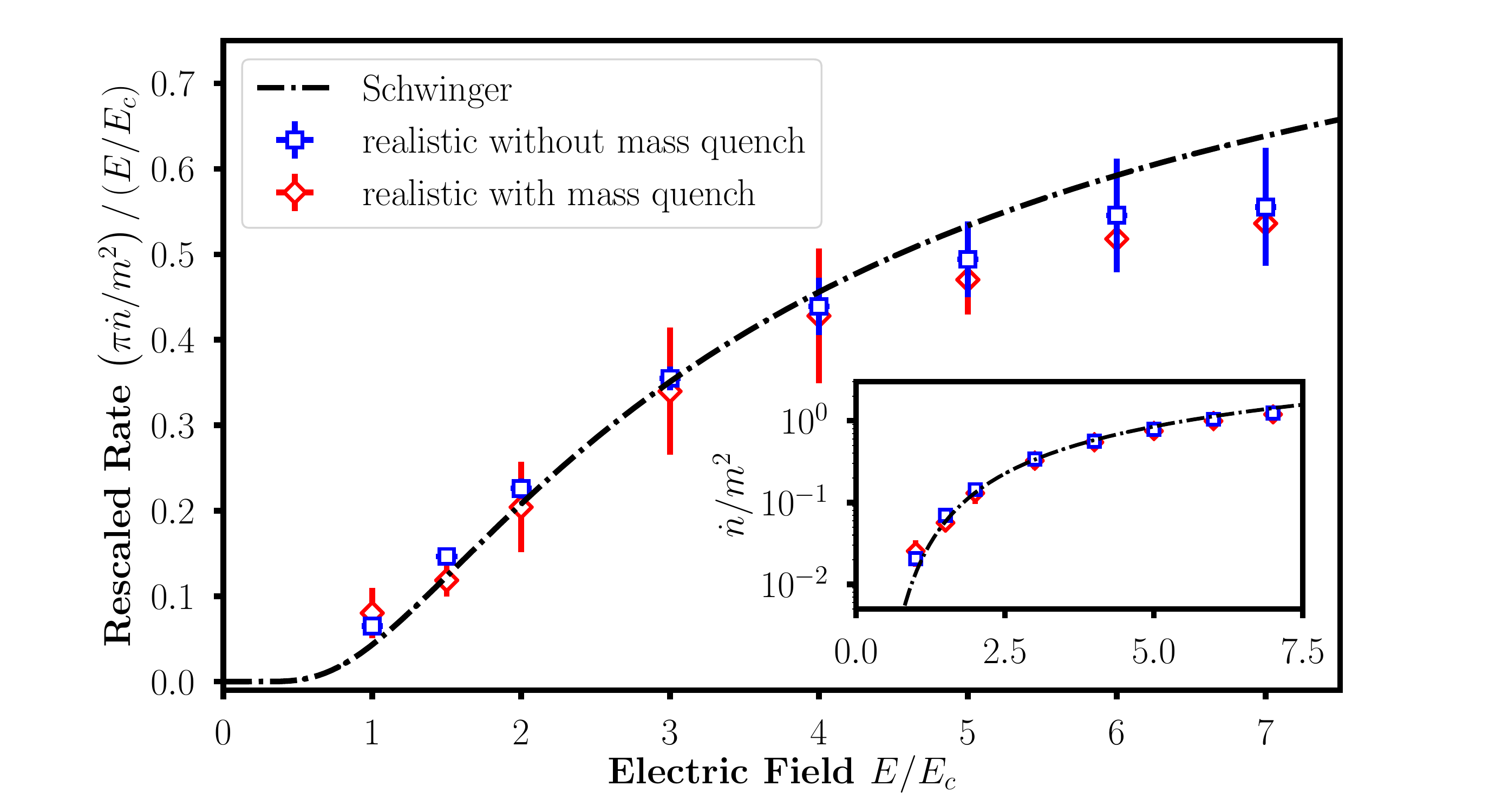}
\caption{\label{rate_E}
	The proposed quantum simulator can quantitatively predict the non-perturbative behavior of the particle-production rate due to Schwinger \cite{schwinger1951gauge}. 
	Blue squares (red diamonds) are extracted from linear fits to results of our benchmark simulations without (with) initial mass quench. The rightmost squares corresponds to the set of realistic parameters used in Fig.~\ref{simulation2} (and \ref{mass_quench}) and the others only differ in the initial electric field. The black dashed-dotted line is the analytic prediction. For the parameters chosen in this work, the results for both initialization procedures agree within the estimated experimental accuracy (see appendix \ref{app:C}).
	Main panel: particle-production rate, rescaled by the linear factor $E/\left(\pi E_c\right)$.  
	Inset: full rate in doubly-logarithmic scale.}
	}
\end{figure}

Even with lattice artifacts, the dependence of the particle production rate on the initial value of the electric field is highly non-trivial. 
In Fig.~\ref{rate_E}, we present the numerically extracted particle production rate after dividing out the `trivial' linear dependence on $E$, for both initialization procedures outlined above and compare it to the analytical prediction \eqref{eq:SchwingerRate}. 
Our simulations indicate that the proposed implementation can reproduce quantitatively the non-perturbative suppression of particle production for weak fields $E < \pi E_c$. 
For larger fields $E > \pi E_c$, we furthermore observe the expected saturation of the rescaled rate, though quantitative agreement is not achieved due to deviations from the continuum limit. Nevertheless, the full rate, plotted in the inset of Fig.~\ref{rate_E} is remarkably close to the analytic prediction over two orders of magnitude for our simulations both with and without initial mass quench. (The excellent agreement for weak fields should be taken with a grain of salt since the rate will be very challenging to extract in a realistic experiment in this region.) For a detailed description of the fitting procedure to obtain the rate, the experimental accessibility of the weak-field regime, and the simulations with initial mass quench we refer to the appendix \ref{app:C}. 
As these results indicate, non-trivial effects of lattice gauge theories are within reach of current optical-lattice technology.

\section{Discussion and conclusions}\label{sec6}
To summarize, exploiting the freedom in discretizing relativistic fermions on a lattice opens up a hitherto unexplored possibility for optimizing quantum simulators in view of experimental implementations. We have exemplified this for the Wilson formulation of lattice fermions, where it enables an elegant implementation of 1+1D QED. 
Our numerical benchmark calculations indicate that available experimental resources may access the Schwinger mechanism of particle-anti-particle production, and in particular extract the non-perturbative onset of the production rate. The performance of the quantum simulator may even be further improved by resorting to mixtures with enhanced spin-changing collisions, such as sodium--potassium.

As a final aim of a cold-atom quantum simulator, it stands to advance into
parameter regimes that are not accessible to computer simulations. For the
optimized parameters in the experimental implementation proposed in this
work, the dimensionless coupling $e/m$ and the lattice spacing $am$ cannot
be tuned independently. Rather, as one is increased, the other one has to be
decreased and vice versa, which can be exploited as follows. First, one can
use the particle-production rate in the regime of small $e/m$ (and thus
large $am$) to benchmark the quantum simulator using numerical methods. The
experiment may then proceed into the relevant regime of strong coupling,
which for the optimized parameters permits us also to reduce the lattice
spacing. This regime of strong coupling is not accessible within current
numerical simulations such as the functional integral method employed here.
The proposed experiment involves $20 \times 2 = 40$ fermionic degrees of
freedom interacting with $20$ BECs of $\mathcal{O}(1000)$ atoms each,
resulting in a Hilbert space size that is also beyond the capability of
exact simulation methods.

Our work also opens an interesting pathway for future theoretical investigations. 
By casting Wilson fermions in the compact form of the Hamiltonian in Eq.~\eqref{1d_Wilson_Hamiltonian}, the free lattice theory becomes manifestly equivalent to the Su-Schrieffer-Heeger model, one of the simplest models displaying non-trivial topological properties \cite{su1979solitons}. 
Indeed, the topological properties of free Wilson fermions have been considered in Ref.~\cite{bermudez2010wilson} and, very recently, in Ref.~\cite{kuno2018generalized}. 
It will be exciting to study these topological properties in the context of interacting gauge theories.

\begin{acknowledgments}
We thank Valentin Kasper for very helpful discussions, Alexander Mil and Fabián Olivares for input on the experimental implementation, and Martin Gärttner for discussions and comments on the manuscript.
This work is part of and supported by the DFG Collaborative Research Centre "SFB 1225 (ISOQUANT)", the ERC Advanced Grant "AtomicGaugeSimulator" (Project-ID 339220), the ERC Advanced
Grant "EntangleGen" (Project-ID 694561), the DFG (Project-ID 377616843), and
the Innovation Fund “Frontier” of Heidelberg University (“Simulating
high-energy physics in small atomic systems”, project-ID: ZUK 49/Ü 5.2.170) 
\end{acknowledgments}

\appendix

\section{Details of the experimental implementation \label{app:A}}
As outlined in the main text, we propose to implement \eqref{cold_atom_H} with a mixture of fermions and bosons trapped in an optical lattice. 
In the following, we discuss some details and subtleties that arise in this implementation. 
Though we keep the discussion general, when giving numerical estimates we assume a mixture of fermionic ${}^6 \text{Li}$ and bosonic ${}^{23} \text{Na}$. 
For the most part, we assume transverse degrees of freedom to be frozen out and consider the mixture to be effectively one-dimensional. 

\subsection{Single-particle Hamiltonian}
In accordance with the structure of Wilson fermions (see Fig.~\ref{Wilson-vs-staggered}), we propose to employ a tilted optical lattice of the form 
\begin{align}\label{optical_potential}
V_{\chi}^{(\text{lat})}(x) = \mathcal{F}_{\chi} x + \begin{cases} V_\chi \cos^2 \left( \frac{\pi x}{a_\text{lat}} \right) \, , & \chi = B \\ V_\chi \sin^2\left( \frac{\pi x}{a_\text{lat}} \right) \, , & \chi = F \end{cases} \; ,
\end{align}
where the species index $\chi=F,B$ refers to either fermions on the lattice sites or bosons on the links.
The depth $V_{\chi}$, the spacing $a_\text{lat}$, and the tilt strength $\mathcal{F}_{\chi}$ of the optical lattice may be tuned independently. Additionally, we apply a constant magnetic field $\mathcal{B}$ perpendicular to the $x$-direction, which we assume to give rise to a linear Zeeman shift,
\begin{align}
V_{\chi,s}^{(\text{Z})} = -m_{\chi,s} g^{(\mathcal{B})}_{\chi,s} \mu_\mathcal{B} \mathcal{B} \; .
\end{align}
Here, $s= \uparrow, \downarrow$ denotes the selected two hyperfine states for each species to which the relevant dynamics of the system is restricted; $m_{\chi,s}$ is the corresponding magnetic quantum number, $\mu_\mathcal{B}$ denotes the Bohr magneton, and $g^{(\mathcal{B})}_{\chi,s}$ is the Landé g-factor. Neglecting interactions for a moment, the quadratic part of the full Hamiltonian is given by
\begin{align}\label{H0full}
H_0 &= \sum_{\chi,s} \int dx \; \chi_s^\dagger(x) \mathcal{H}_0(\chi,s) \chi_s(x)\; \\ 
\mathcal{H}_0(\chi,s) &=  -\frac{\hbar^2 \partial_x^2}{2M_{\chi}} + V_{\chi}^{(\text{lat})}(x) + V_{\chi,s}^{(\text{Z})} \; , \label{H_0}
\end{align}
where $M_{\chi}$ is the atomic mass and we assumed the lattice potential to be species-dependent, but the same for different hyperfine states. The fields, which we denote by $\chi_s$, obey canonical commutation or anti-commutation relations according to their statistics, i.e.\ 
\begin{align}
\left[\chi_s(x), \chi^\dagger_r(y)\right]_{\zeta(\chi)} = \delta(x - y) \; ,
\end{align}
where we abbreviate $\left[X,Y\right]_\pm = XY \pm YX$ with $\zeta(B) = -$ and $\zeta(F) = +$.

\subsection{Suppression of direct tunneling in a tilted lattice}
Additionally to its matching the natural structure of Wilson fermions, we employ the tilt to suppress direct tunneling of the fermions. This is crucial to ensure gauge invariance, because the fermions must only hop between different lattice sites due to interactions with the bosons. In the untilted case ($\mathcal{F}_\chi=0$), the one-particle Bloch waves for the potential in Eq.~\eqref{H_0} without external magnetic field ($\mathcal{B}=0$) are Mathieu functions. In this case, the ground band has the dispersion relation 
\begin{align}
\epsilon_{\chi}(k) = \frac{\omega_{\chi}}{2} - 2J_{\chi} \cos(ka_\text{lat}) \; , &&k \in \left[-\frac{\pi}{a_\text{lat}}, \frac{\pi}{a_\text{lat}}\right)\,,
\end{align}
with the mean energy $\omega_{\chi} = 2 \sqrt{V_{\chi} E_{\mathrm{rec},\chi}}$. Here $E_{\mathrm{rec},\chi} = \hbar^2 /\left[2M_{\chi} (a_\text{lat}/\pi)^2\right]$ is the recoil energy and the ratio $\xi_{\chi} = 2\sqrt{V_{\chi}/E_{\mathrm{rec},\chi}}$ controls the lattice depth. In the limit of a deep lattice $\left(\xi_{\chi} \gg 1 \right)$ the hopping element is given by
\begin{align}
J_{\chi} = \sqrt{\frac{2}{\pi}} E_{\mathrm{rec},\chi} \left(\xi_{\chi}\right)^{3/2} e^{-\xi_{\chi}}\left[ 1+ \mathcal{O}(1/\xi_{\chi})\right],
\end{align}
which can be obtained exactly from analytic properties of the Mathieu functions.

We can suppress direct tunneling for both species by choosing a sufficiently strong tilt, i.e.
\begin{align}\label{supp_tunnel}
a_\text{lat}\mathcal{F}_{\chi} \gg J_{\chi} \, . 
\end{align} 
In the presence of the tilt, the states in the ground band are modified into resonances of a Wannier-Stark ladder (see e.g.\ Ref.~\cite{gluck2002wannier}).
Including non-vanishing $\mathcal{B}\neq 0$, the energy levels are
\begin{align}
\mathcal{E}_{F,s}(l) &= \frac{1}{2} \omega_{F} + l a_\text{lat} \mathcal{F}_{F} -m_{F,s} g_{F,s}^{(\mathcal{B})} \mu_\mathcal{B} \mathcal{B} \; , \\
\mathcal{E}_{B,s}(l) &= \frac{1}{2} \omega_{B} + \left(l + \frac{1}{2} \right) a_\text{lat} \mathcal{F}_{B} -m_{B,s} g_{B,s}^{(\mathcal{B})} \mu_\mathcal{B} \mathcal{B} \; .
\end{align}

The states of the Wannier-Stark ladder have a finite lifetime which can be estimated from the decay rate $\Gamma_{\chi}$ due to Landau-Zener transitions \cite{gluck2002wannier}, 
\begin{align}
\Gamma_{\chi} = \frac{a_\text{lat}\mathcal{F}_{\chi}}{2\pi \hbar} \exp \left( - \frac{\pi^2 \Delta_{\chi}^2}{8 E_{\chi} a_\text{lat} \mathcal{F}_{\chi}} \right) \le \frac{a_\text{lat}\mathcal{F}_{\chi}}{2\pi \hbar} \; ,
\end{align}
where $\Delta_{\chi} \approx \omega_{\chi}$ is the gap between the ground band and the first excited band. 
This lifetime is one of the relevant experimental restrictions.

\subsection{Choice of the magnetic field}
Under the above condition \eqref{supp_tunnel}, we may neglect direct tunneling. Then the quadratic part \eqref{H0full} of the full Hamiltonian amounts to a Wannier-Stark ladder of long-lived resonances. These are coupled by a correlated hopping of the fermions and bosons as in Ref.~\cite{kasper2017implementing}, which induces the gauge-invariant matter--gauge-field interaction. 
For this purpose, the spin-changing collisions for the chosen hyperfine levels $s= \uparrow, \downarrow$ need to be tuned into resonance, i.e.\ we demand 
\begin{align}
\mathcal{E}_{F\uparrow} \left(l\right) - \mathcal{E}_{F\downarrow}\left(l+1\right) \overset{!}{=} \mathcal{E}_{B\uparrow}\left(l\right) - \mathcal{E}_{B\downarrow}\left(l\right) \;   \qquad \forall l \; .
\end{align}
If the two components $\uparrow, \downarrow$ are chosen from the same hyperfine manifold for each species, the g-factors are spin-independent, $g^{(\mathcal{B})}_{\chi,s} = g^{(\mathcal{B})}_\chi$ . Then the resonance conditions can be rewritten as
\begin{align}
\frac{a_\text{lat}\mathcal{F}_F}{\mu_\mathcal{B} \mathcal{B}} \overset{!}{=} g^{(\mathcal{B})}_B \Delta_B^{(\mathcal{B})} - g^{(\mathcal{B})}_F \Delta_F^{(\mathcal{B})} \;  ,
\end{align}
where we introduced the abbreviation $\Delta_\chi^{(\mathcal{B})} = m_{\chi\uparrow} - m_{\chi\downarrow}$. 

For a mixture of bosonic ${}^{23}$Na and fermionic ${}^6$Li, respectively, we choose the following levels from the ground hyperfine manifold: 
\begin{align}
m_{B\uparrow} = 0 \; , \quad m_{B\downarrow} = -1 \, , \quad m_{F\uparrow} = \frac{1}{2} \; , \quad m_{F\downarrow} = -\frac{1}{2} \; .
\end{align}
The corresponding Landé factors are $g^{(\mathcal{B})}_F = -\frac{2}{3}$ and $g^{(\mathcal{B})}_B = -\frac{1}{2}$.

\subsection{Effective interaction constants from overlap integrals in the tilted lattice} 
In order to match the coefficients of Hamiltonian \eqref{cold_atom_H} with experimental parameters, we need to calculate overlap integrals involving Wannier-Stark functions $\Psi_{l,\chi}(x)$. In our case, they are built from the Wannier functions $\psi_{l,\chi}(x)$ located at lattice sites (respectively links) $l=1\dots N$ of the untilted lattice. The Wannier-Stark functions can be written as superpositions 
\begin{align}\label{WS_functions}
\Psi_{l,\chi}(x) = \sum_m \mathcal{J}_{m-l}\left( \frac{2 J_{\chi}}{a_\text{lat} F_{\chi}}\right) \psi_{m,\chi}(x) \; ,
\end{align}
where $\mathcal{J}_m(\dots)$ denote Bessel functions of the first kind. We only consider the ground band here. For sufficiently deep lattices, to estimate the relevant overlap integrals, we may approximate the Wannier functions appearing in the series \eqref{WS_functions} as harmonic oscillator eigenfunctions, 
\begin{align}
\label{eq:HOapproximation}
\psi_{l,\chi}(x) = \left(\pi a_{\chi}^\text{HO} \right)^{-1/4} \exp \left[-\frac{1}{2} \left(\frac{x-x_{l,\chi}}{a_{\chi}^\text{HO}} \right)^2\right] \; , 
\end{align}
where $x_{l,\chi}$ denote the minima of the tilted potentials and the harmonic oscillator lengths should be determined from a Taylor expansion of the tilted potentials around their minima. The minima are shifted from the untilted case to the positions 
\begin{align}
&x_{l,B} = \left(l+\frac{1}{2} \right)a_\text{lat} - \delta_B \, , \qquad x_{l,F} = la_\text{lat} - \delta_F \, ,\\ 
&\delta_\chi = \frac{a_\text{lat}}{2\pi} \arcsin \left( \frac{a_\text{lat}\mathcal{F}_\chi}{\pi V_\chi} \right) \approx  \frac{a_\text{lat}}{2\pi} \frac{a_\text{lat}\mathcal{F}_\chi}{\pi V_\chi} \, .
\end{align}
The oscillator length, $a_{\chi}^\text{HO} = \sqrt{\hbar /\left(m_{\chi} \omega_{\chi}^\text{HO}\right)}$, is determined from the condition 
\begin{equation}
m_{\chi} \left(\omega_{\chi}^\text{HO}\right)^2 \overset{!}{=} 2V_\chi \left( \frac{\pi}{a_\text{lat}}\right)^2 \sqrt{1- \left( \frac{a_\text{lat}\mathcal{F}_\chi}{\pi V_\chi} \right)^2}\,.
\end{equation}

Using the approximation \eqref{eq:HOapproximation}, we can calculate the effective interaction constants that enter the quantum simulation. For clarity, we first present estimates based on single-particle wavefunctions. 
For the bosonic condensates, we subsequently use a more appropriate estimate to take their high occupation into account. 

The effective boson-fermion interspecies interaction generating the matter-field hopping is given by the 3D overlap integral 
\begin{align}
\label{eq:g_BF}
g_{BF} &= \frac{g^{(3D)}_{BF}}{2} \int dy \int dz \, \left|\Phi_B(y,z)\right|^2 \; \left|\Phi_F(y,z)\right|^2 \nonumber\\&\times \int dx \,  \left(\Psi_{l+1,F}(x)\right)^* \left| \Psi_{l,B}(x) \right|^2 \Psi_{l,F}(x) \; ,
\end{align}
where $g^{(3D)}_{BF} = \left(\sqrt{2}/3\right)2\pi \hbar^2 a_{B F}/ M_\text{red}$ is the relevant three-dimensional interaction constant with scattering length $a_{BF} \approx 0.9 a_0$
and reduced mass $M_\text{red} = M_F M_B/\left(M_F + M_B\right)$. The Clebsch-Gordon coefficient $\sqrt{2}/3$ is the same as for the previous proposal \cite{kasper2017implementing}.  For simplicity, we assume a symmetrically harmonic transverse confinement with the same trapping frequency $\omega_\perp$ for $B$ and $F$, i.e.
\begin{align}
\Phi_\chi(y,z) =  \left(\pi a_{\perp,\chi}^2 \right)^{-1/2} \exp \left( - \frac{y^2 + z^2}{2a_{\perp,\chi}^2}\right) \; 
\end{align}
and $a_{\perp,\chi} = \sqrt{\hbar/\left(M_\chi \omega_\perp\right)} \; .$

Furthermore, we need to calculate the effective bosonic intraspecies interaction
\begin{align}
\label{eq:gB}
g_B = \frac{g_B^{(3D)}}{2}\int dy \int dz \; \left|\Phi_B(y,z)\right|^4 \int dx \; \left|\Psi_{l,B}(x)\right|^4 \; ,
\end{align}
where $g_B^{(3D)} = \left(1/6\right)4\pi \hbar^2 a_{B}/M_B$ is the relevant interaction strength with $a_B \approx 5 a_0$. For the Clebsch-Gordon coefficient, we again refer to \cite{kasper2017implementing}.
At the large boson occupation numbers that we are interested in, the effective interaction constant is modified due to deviations of the bosonic wave-function from the harmonic-oscillator shape.  
In a first approximation, we may treat this effect in a Thomas-Fermi limit, i.e.\ we consider the bosonic mean-field wavefunction as
\begin{align}
\Phi_\text{TF}(x) = \sqrt{\frac{\tilde{\mu}}{\tilde{g}}\left[ 1- \frac{x^2}{x_\text{TF}^2} \right]} \Theta \left( |x| - x_\text{TF}\right)
\end{align}
with the Thomas-Fermi radius $x_\text{TF} = \sqrt{2\tilde{\mu}/\left[M_B \left(\omega_B^\text{HO}\right)^2\right]}$ and the effective chemical potential $\tilde{\mu}$ is determined by the number of bosons on each link as $N_B = 4 x_\text{TF} \tilde{\mu}/\left(3\tilde{g}\right)$.
Thus, in terms of the experimental parameters,
\begin{align}
x_\text{TF} = \left\lbrace \frac{4 V_B}{3 \tilde{g} N_B} \left( \frac{\pi}{a} \right)^2 \sqrt{1- \left( \frac{a_{\mathrm{lat}} \mathcal{F}_B}{\pi V_B} \right)^2} \right\rbrace^{-1/3} \, ,
\end{align}
where $\tilde{g}$ is the effective 1D interbosonic coupling given by
\begin{align}
\tilde{g} = \frac{g_B^{(3D)}}{2}\int dy \int dz \; \left|\Phi_B(y,z)\right|^4 \; .
\end{align}
Due to the vastly different trapping frequencies (few $\SI{}{\hertz}$ compared to several $\SI{}{\kilo\hertz}$, see below), it is justified to treat the bosons in the radial direction in the harmonic approximation, while taking into account the Thomas-Fermi profile in the longitudinal direction. The results of the main section are obtained with the effective interaction constants calculated within this approximation. 

We note that for a general choice of fermion species, on-site density-density interactions can yield unwanted effects. However, for the present choice of Lithium, these are absent due to a zero of the relevant scattering length at low magnetic fields \cite{houbiers1998elastic}.

\subsection{Three-body losses}
One of the leading limitations of our proposal is the instability of the local BECs due to three-body collisions. On a mean-field level, these three-body losses may be modelled as
\begin{align}\label{3body_loss}
\frac{\dot{N}_{B}}{N_{B}} = - \frac{K_3}{N_{B}} \int d^3x\, \left| \psi_B(x,y,z) \right|^6\,,
\end{align}
with the total particle number per link $N_{B} = \int d^3x\, \left| \psi_B(x,y,z) \right|^2$ and the specied-dependent constant $K_3$, commonly referred to as the three-body loss rate coefficient. We estimate the typical time-scale for three-body losses, $T_3$, as the inverse of the right-hand side of equation \eqref{3body_loss}. Demanding this time to be much larger than any other time scale in the experiment sets a limit on the boson particle number $N_{B}$ per link. 

\section{Choice of experimental parameters\label{app:B}}
In the implementation proposed above, several imperfections arise that limit accessible time scales. In this section, we discuss the main limitations and an optimization procedure to maximize the time simulatable in the experiment. 

\subsection{Experimental limitations\label{sec:experimentalLimitations}}
A first limitation appears through the three-body loss time of the bosonic condensates, $T_3$. 
Second, the localised Wannier-Stark states in the tilted periodic potential have a finite life-time. 
For the choice of a mixture between ${}^6 \text{Li}$ and ${}^{23} \text{Na}$, due to the smaller mass of the fermionic atoms the limiting factor will be the fermion life time, which we denote by $T_{LZ}$. 
Third, the proposed lattice QED implementation will have a finite lattice spacing  $a$. Deviations from the continuum limit will appear at a time scale $T_{\mathrm{lat}}$, which is defined below.

One can balance between these three effects by adapting the following experimental parameters:
\begin{enumerate}
\item The number of bosons per link $N_{B}$,
\item the number of lattice sites $N$,
\item the optical lattice spacing $a_\text{lat}$,
\item the lattice depth $V_\chi$,
\item the radial confinement $\omega_{\perp, \chi}$,
\item the tilt of the lattice $a_\text{lat} \mathcal{F}_{\chi}$.
\end{enumerate}
Allowing different parameters for the bosons and fermions, the three time-scales depend on $9$ different experimental parameters. We are looking for a set of parameters for which the minimum of $\left(T_3, T_{LZ}, T_{\mathrm{lat}}\right)$ becomes maximal. This optimization is constrained by experimental restrictions, e.g.\ the achievable number of lattice sites. 

In addition, each parameter set corresponds to two dimensionless parameters on the lattice QED side, namely the lattice spacing $am$ and the coupling constant $e/m$ in units of the fermion mass $m$. We wish to obtain relevant results for continuum QED and thus seek small values of $am \lesssim 1$. On the other hand, we want to perform benchmarking simulations employing an approximation which relies on $e/m < 1$. These different desiderata generate a rather complicated optimization problem.

\subsection{Simplified optimization procedure}

Relying on physical intuition, we simplify the problem by considering the limiting time-scales as a function of $a_\text{lat}$ in the range $5-10 \SI{}{\micro \meter}$. Since one of the main complications consists of realizing strong spin-changing collisions, we choose rather shallow lattices, $V_{B} = 2 E_{\mathrm{rec},B}$ and $V_{F} = E_{\mathrm{rec},F}$ and fix the tilts $a_{\mathrm{lat}}\mathcal{F}_\chi = 10 J_\chi$ such that direct tunneling in the shallow lattice is suppressed. 
We further fix a strong radial confinement $\omega_{\perp, \chi} = 2 \pi \times  \SI{10}{\kilo \hertz}$ that renders the system effectively one-dimensional. Now we can tune $N_{B}$, such that $T_3$ and $T_{LZ}$ lie approximately in the same range. We choose $N_{B} = 3000$, i.e.\ $\ell = 1500$, which is sufficiently large for the quantum link regularisation to approximate QED. Finally, the total number of lattice sites is irrelevant for the determination of three time-scales, but enters as the IR-cutoff after having determined the simulation parameter $am$. We will come back to this point later.

We can now calculate the effective bosonic interaction constant $(g_B)$ and the effective interaction constant for correlated hopping between the fermions and bosons $(g_{BF})$ as functions of $a_\text{lat}$. Finally, we may adjust the local oscillation frequency between the two fermionic states $(\Omega)$. Then, we can make the connection to lattice QED according to Eqs.~\eqref{omega}, \eqref{g_boson}, and \eqref{g_SCC}. On the QED side, we choose to measure energies and times in units of $m$, which in experimental parameters corresponds to $m \leftrightarrow  \Omega - 2 \sqrt{\ell (\ell +1)} g_B$. 
As a bound for the validity of the lattice simulation, we take $T_{\mathrm{lat}}m = 2\pi/\left(am\right)$, which is the time when the momentum $p_{\mathrm{lat}} = eET_{\mathrm{lat}}$ of particles accelerated by a constant electric field $E=m^2/e$ reaches the cutoff $\sim 2\pi/a$. 
Within this framework, we consider a two-step procedure choosing first $a_\text{lat}$ and subsequently $\Omega$ in order to optimize the functions $T_3(a_\text{lat})$, $T_{LZ}(a_\text{lat})$ and $T_{\mathrm{lat}}(a_\text{lat}, \Omega)$, $am (a_\text{lat}, \Omega)$, $e/m (a_\text{lat}, \Omega)$.

\begin{figure}\centering{
\includegraphics[scale=0.32]{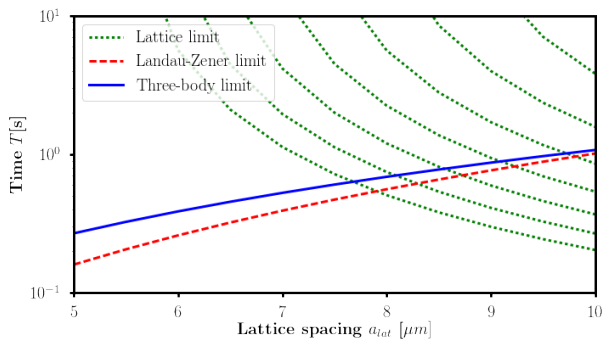}}\caption{\label{time-scales} The limiting time-scales $T_3(a_\text{lat})$ (blue, solid), $T_{LZ}(a_\text{lat})$ (red, dashed), and $T_{\mathrm{lat}}(a_\text{lat},\Omega)$ (green, dotted; top to bottom for $\Omega = 2\pi \times  4-7\, \SI{}{\hertz}$ in steps of $0.5\, \SI{}{\hertz}$). An optimal choice of experimental parameters maximizes the minimum of these three time scales, conditioned on the desired set of simulation parameters, see \ref{QEDpars}. Note the logarithmic scale of the ordinate.}
\end{figure}
\begin{figure}
\centering{
\includegraphics[scale=0.32]{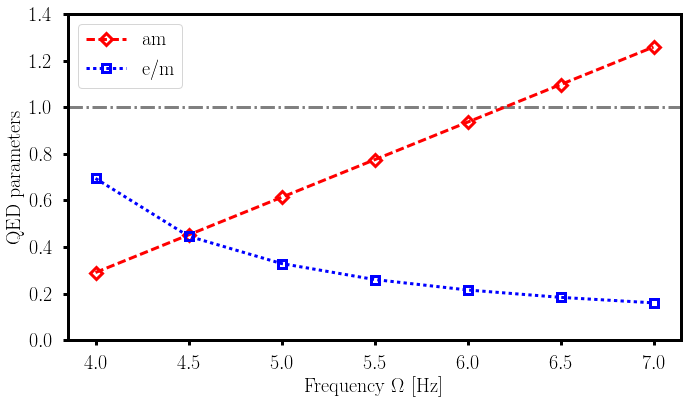}}\caption{\label{QEDpars} Dimensionless parameters entering the lattice QED simulation, $am \, (\Omega)$ (red diamonds, dashed), and $e/m \, (\Omega)$ (blue squares, dotted). The horizontal grey dashed-dotted line indicates the limiting unity. The inverse behaviour favours strong coupling at small lattice spacing. }
\end{figure}

Figure \ref{time-scales} shows the limiting time-scales for different values of $\Omega$ in the range of $ 2\pi \times (4-7)\SI{}{\hertz}$. As this plot shows, large values of $a_\text{lat}$ are favourable if we adjust $\Omega$ accordingly. Consequently, we choose a rather large $a_\text{lat} = \SI{10}{\micro \meter}$ and plot the remaining $\Omega$ dependence of the lattice QED parameters in Fig.~\ref{QEDpars}. The observed converse behaviour of $am$ and $e/m$ restrains us from choosing an arbitrarily small lattice spacing $a$ for the benchmarking simulations. Note that this should not be seen as a problem of the implementation, but rather getting close to the continuum limit means studying the strong-coupling regime of QED. This regime is notoriously difficult for numerical simulations and thus a non-trivial target for quantum simulation.  
For benchmarking, however, we choose the following two possibilities:
\begin{align}
\Omega = 2\pi \times \SI{4.5}{\hertz}: &&am = 0.45 \;, \quad e/m = 0.45\\
\Omega = 2\pi \times \SI{6}{\hertz}: && am = 0.94 \;, \quad e/m = 0.22
\end{align}
Finally, we choose the number of lattice sites to be $N = 20$, which corresponds to a reasonable size of the optical lattice. The corresponding IR-cutoff on the QED side in units of the fermion mass is given by $1/\left(N \times am\right) \sim \mathcal{O}(0.1)$. Thus, we should be able to resolve sufficiently many modes at small momenta to observe the phenomenon of Schwinger pair production. 

\section{Details of the numerical simulation\label{app:C}}
For our numerical simulations, we employ a functional-integral approach that was also used to benchmark a previous implementation with staggered fermions \cite{kasper2016schwinger,kasper2017implementing}. The main idea of this method is to map the full quantum theory onto a classical-statistical ensemble, which is achieved by a semi-classical expansion around the intial state. Observables are then obtained by solving classical equations of motion and sampling over fluctuating initial values. This results in a non-perturbative approximation of the quantum dynamics, which is valid for sufficiently large electric fields $E$ and weak coupling $e/m$. In the context of cold atomic gases, the method is related to the well-known truncated Wigner approximation and can be derived by integrating out the fermionic degrees of freedom. The validity to benchmark the proposed quantum simulator relies on the fact that the dynamics is dominated by the Bose condensates and that direct interactions between the fermions are absent.

\begin{widetext}
\subsection{Classical equations of motion}
In order to calculate the equations of motions for the classical-statistical approach to Bose-Fermi mixtures, we consider the Weyl symbol of the cold atom Hamiltonian \eqref{cold_atom_H}, which is given by
\begin{align}
H &= \frac{ae^2}{4} \sum_n \left(|b_n|^4 + |d_n|^4\right)
+ \left(m+ \frac{1}{a}\right)\sum_n \left(\psi_{n,1}^\dagger \psi_{n,2} + \psi_{n,2}^\dagger \psi_{n,1}\right)
 + \frac{1}{a \sqrt{\ell\left(\ell + 1\right)}} \sum_n \left( \psi_{n-1,1}^\dagger b_{n-1}^* d_{n-1} \psi_{n,2} + \text{h.c.}\right) \; .
\end{align}
In this expression, $b_n,d_n$ are c-numbers and $\psi_{n,\alpha}$ are fermionic operators at site $n$ and hyperfine state $\alpha$. Thus, we can decompose the Hamiltonian to $H = H_B + H_F$, with the pure c-number part
$H_B = \frac{ae^2}{4} \sum_n \left(|b_n|^4 + |d_n|^4\right)$ and the fermionic part $H_F = \sum_{mn, \alpha \beta}\psi_{m,\alpha}^\dagger \left(h_F\right)_{mn}^{\alpha \beta} \psi_{n, \beta}$, where we abbreviated
\begin{align}
\left(h_F\right)_{mn}^{\alpha \beta} &= \left(m+ \frac{1}{a}\right) \delta_{mn}\sigma_x^{\alpha \beta} + \frac{1}{a \sqrt{\ell \left(\ell + 1\right)}} \left( b_m^* d_m \delta_{m, n-1} \sigma_+^{\alpha \beta} + d_{m-1}^* b_{m-1} \delta_{m-1, n} \sigma_-^{\alpha \beta}\right)
\end{align}
with the Pauli matrices
\begin{align}
\sigma_x = \begin{pmatrix}
0 & 1 \\ 1 & 0
\end{pmatrix} \; , && \sigma_y = \begin{pmatrix}
0 & -i \\ i & 0  
\end{pmatrix} \; , && \sigma_\pm = \frac{1}{2}\left(\sigma_x \pm i \sigma_y \right) \; .
\end{align}
In terms of the equal-time two-point function $D^{\alpha \beta}_{mn} = \left\langle \psi_{m,\alpha}^\dagger \psi_{n,\beta} \right\rangle$, where $\langle \dots \rangle$ denotes a quantum expectation value, the explicit equations of motion for the classical-statistical theory are derived from
\begin{align}
i \partial_t c_n = \frac{\partial H_A}{\partial c_n^*} + \text{Tr} \left[\frac{\partial \left(h_F\right)}{\partial c_n^*} D\right] \; ,  &&
i \partial_t D_{mn}^{\alpha \beta} = \left[h_F, D\right]_{mn}^{\alpha \beta}\; ,
\end{align}
where $c \in \lbrace b,d \rbrace$ and the trace runs over spatial and fermionic species indices $m,n$ and $\alpha, \beta$, respectively. Rewriting $D_{mn}^{\alpha \beta} = \frac{1}{2}\delta_{mn}\delta^{\alpha \beta} -F_{nm}^{\beta \alpha}$ with the statistical propagator $F_{mn}^{\alpha \beta} =\frac{1}{2} \left\langle \left[ \psi_{m,\alpha} , \psi_{n,\beta}^\dagger \right] \right\rangle$, the full set of equations of motion can be reduced to
\begin{subequations}
\begin{align}
i \partial_t b_n &= \frac{ae^2}{2} \left| b_n \right|^2 b_n - \frac{d_n}{a \sqrt{\ell(\ell +1)}} \left[F_{n+1,n}^{21}\right]^* \; , \\
i \partial_t d_n &= \frac{ae^2}{2} \left| d_n \right|^2 d_n - \frac{b_n}{a \sqrt{\ell(\ell +1)}} F_{n+1,n}^{21} \; , \\
i \partial_t F_{nm}^{21} &= \left(m+ \frac{1}{a}\right)\left(F_{nm}^{22} - F_{nm}^{11}\right) +\frac{1}{a\sqrt{\ell(\ell +1)}}\left(b_m^* d_m F_{n,m+1}^{22}-b_{n-1}^* d_{n-1}F_{n-1,m}^{11}\right) \; ,\\
i \partial_t F_{nm}^{11} &= \left(m+ \frac{1}{a}\right)\left(\left[F_{mn}^{21}\right]^* - F_{nm}^{21}\right) + \frac{1}{a\sqrt{\ell(\ell +1)}} \left(b_{m}^* d_{m}  \left[F_{m+1,n}^{21}\right]^* - d_n^* b_nF_{n+1,m}^{21}\right) \;, \\
i \partial_t F_{nm}^{22} &= \left(m+ \frac{1}{a}\right) \left(F_{nm}^{21} - \left[F_{mn}^{21}\right]^*\right) + \frac{1}{a\sqrt{\ell(\ell+1)}}\left(d_{m-1}^* b_{m-1} F_{n,m-1}^{21} - b_{n-1}^*d_{n-1}\left[F_{m,n-1}^{21}\right]^*\right) \; .
\end{align} 
\end{subequations}
Solving these equations numerically for the initial values specified in the next subsection allows us to benchmark our proposed implementation with Wilson fermions. Since the dynamics is dominated by the coherent electric field, a single run with given initial conditions already gives a good approximation.
For the purpose of this paper, we therefore omit the statistical sampling of fluctuating initial values, similar to what was done in Refs.~\cite{kasper2016schwinger,kasper2017implementing}.
\end{widetext}

\subsection{Initial values}
As discussed in the main text, we consider Schwinger pair production as a test of the proposed implementation of cold-atom QED. Accordingly, we initialize a fermionic vacuum state in the presence of a strong electric field that exceeds the critical value $E_c=m^2/e$.
\subsubsection{Gauge sector}
In the gauge sector, we initialise a coherent electric field as a coherent spin state with expectation value $L_{z,n}^{(0)}$. The constraint $2\ell = b_n^\dagger b_n + d_n^\dagger d_n$ is Wigner transformed to the c-number expression $2 \ell = \left|b_n\right|^2 + \left|d_n\right|^2 - 1$, which allows for solving the Weyl symbol of the spin operator, $L_{z,n} = \frac{1}{2}\left(\left|b_n\right|^2 - \left|d_n\right|^2 \right)$, for $b$ or $d$ as $|b_n|^2 = \left(\ell + \frac{1}{2}\right) + \left(L_{z,n}\right)_W \; ,\;\; |d_n|^2 = \left(\ell + \frac{1}{2}\right) - \left(L_{z,n}\right)_W$. Thus, we choose the initial values
\begin{align}
b_n (t_0) = \sqrt{\left(\ell + \frac{1}{2}\right) + L_{z,n}^{(0)}} \; , \\
d_n (t_0) = \sqrt{\left(\ell + \frac{1}{2}\right) - L_{z,n}^{(0)}} \; ,
\end{align}
where $L_{z,n}^{(0)} \in \left[ - \ell, \ell \right]$. We choose a homogeneous initial value $L_{z,n}^{(0)} = E_0/e$ with electric field $E_0 = 7 E_c$ and $E_0 = 3 E_c$, respectively, for the two sets of optimized experimental parameters. These values are related to the initial bosonic imbalance as $\Delta N = 2E_0/e \sim \mathcal{O}(100) \ll \ell$. 
To ensure that three-particle losses are not only irrelevant for the absolute number of $N_{B}$ atoms but also for their relative distribution among the two hyperfine states, we stop the simulations at $\mathcal{O}(\SI{100}{\milli \second})$ instead of $\mathcal{O}(\SI{1}{\second})$. 

\subsubsection{Fermion sector}
We want to initialize the fermion sector in a vacuum state, i.e.\ a ground state of the fermionic part $H_\psi$ without electric fields. To this end, we diagonalize $H_\psi = \sum_k \omega_k \left(a_k^\dagger a_k + c_k^\dagger c_k -1\right)$, by the canonical transformation
\begin{align}
\begin{pmatrix}
a_k \\ c_k^\dagger
\end{pmatrix} = 
\frac{1}{\sqrt{2}}\begin{pmatrix}
\psi_{k,1} +\frac{z_k}{\omega_k}\psi_{k,2}  \\ 
\psi_{k,1} -\frac{z_k}{\omega_k}\psi_{k,2} 
\end{pmatrix} 
\end{align}
in Fourier space, $\psi_{k,\alpha}  = \frac{1}{\sqrt{N}} \sum_{n=0}^{N-1} e^{-2 \pi i n k / N} \psi_{n,\alpha} $, with the dispersion relation $\omega_k = \left| z_k \right|$, where $z_k = m+\frac{1}{a} \left(1+ \exp \left(\frac{2\pi i k}{N}\right) \right)$. The ground state $|\Omega \rangle$ is defined by $a_k|\Omega \rangle = c_k|\Omega \rangle = 0 $, which translates into the following initial conditions for the fermionic propagator in momentum space,
\begin{align}\label{intial_fermion_corr}
F_{kk}^{11}(t_0) = 0 \; , && F_{kk}^{22}(t_0) = 0 \; , && F_{kk}^{21}(t_0) = \frac{\omega_k}{2z_k} \; ,
\end{align}
and $F_{kk'}^{\alpha \beta}(t_0)=0$ for $k \neq k'$. 
These correlators are related to  position space via $F_{kk'}^{\alpha \beta} = \frac{1}{N} \sum_{nm} e^{-\frac{2\pi i}{N} \left(mk' - nk\right)} F_{mn}^{\alpha \beta}$.

\subsubsection{Gauge invariance}
The chosen initial conditions fulfill the Gauß law
\begin{align}
\langle G_n \rangle &=  \frac{1}{2} \left(|b_n|^2 - |d_n|^2\right) \nonumber\\
&\quad- \frac{1}{2} \left(|b_{n-1}|^2 - |d_{n-1}|^2\right) + F_{nn}^{11} + F_{nn}^{22} = 0 \; .
\end{align}
This constraint is satisfied during the time evolution by construction, which can also be verified explicitly by applying the equations of motion.

\subsection{Observables}
We extract the total electric field as
\begin{align}\label{Efield}
\frac{E}{E_c} = \frac{e^2}{m^2}\frac{1}{2N}\sum_n \left(\left|b_n\right|^2 - \left|d_n\right|^2 \right)\,.
\end{align}
The total fermionic particle number density is
\begin{align}
n =  \frac{1}{L}\sum_k \tilde{n}_k \; ,
\end{align}
where $\tilde{n}_k = \langle \tilde{a}_k^\dagger \tilde{a}_k + \tilde{c}_k^\dagger \tilde{c}_k \rangle = \frac{\tilde{\epsilon}_k}{\tilde{\omega}_k} + 1$ and the tilde denotes quantities derived from the instantaneous diagonalization of the fermionic part of the Hamiltonian in the homogeneous background of $b,d$. 
Then 
\begin{align}
\tilde{\epsilon}_k = -  \left( \tilde{z}_k F_{kk}^{21} + \left[\tilde{z}_k F_{kk}^{21}\right]^* \right) \; 
\end{align}
with $\tilde{z}_k (b,d) = M+\frac{1}{a} \left(1+ d^* b \exp \left(\frac{2\pi i k}{N}\right) \right)$ and $\tilde{\omega}_k = |\tilde{z}_k|$.

\begin{figure}
\centering{\includegraphics[scale=0.35]{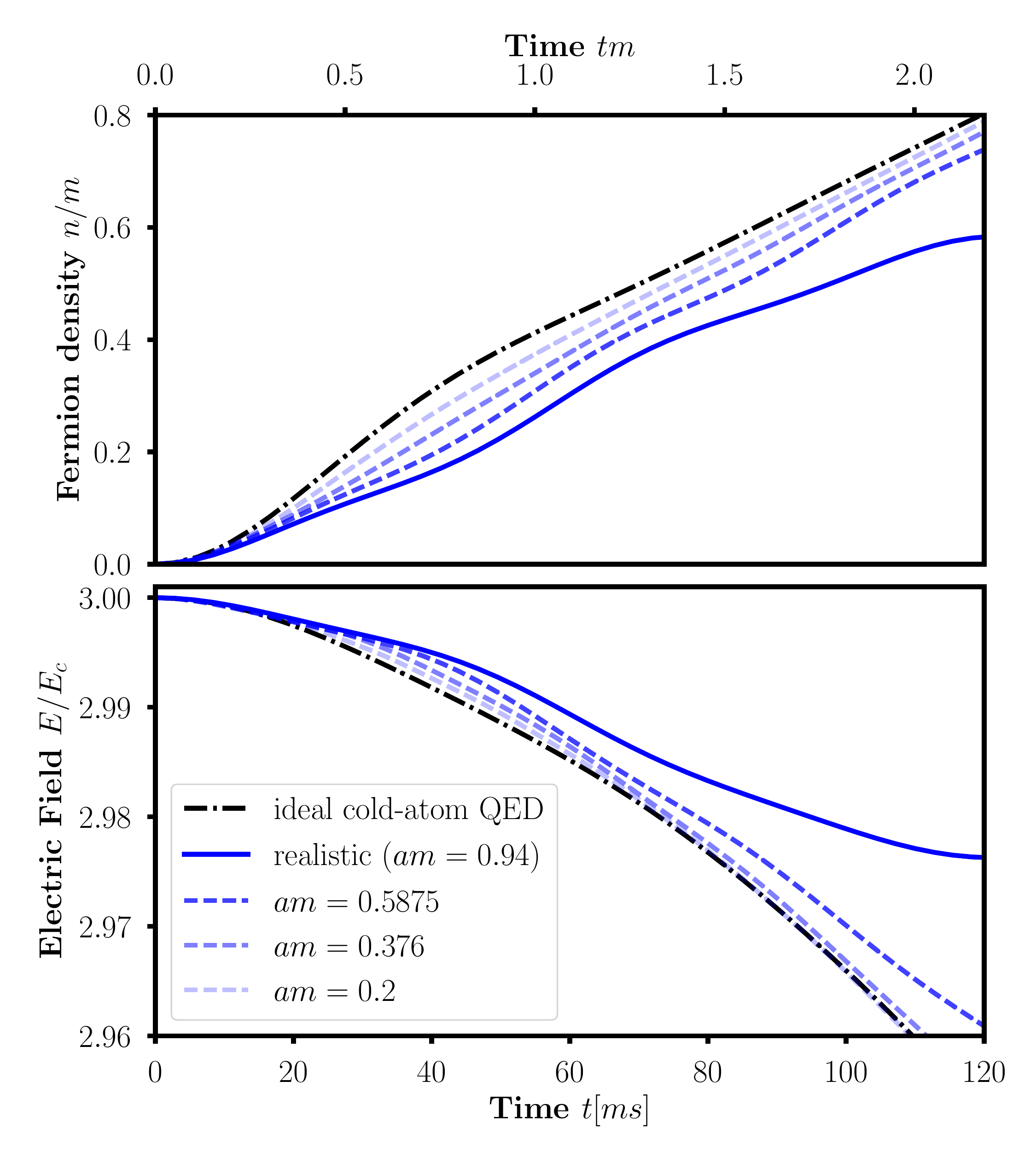}
\caption{\label{simulation1}Benchmarking simulations at $e/m = 0.22$. For the experimentally relevant value of $am = 0.94$,  stronger deviations from the continuum limit can be observed than in Fig.~\ref{simulation2}. 
Nevertheless, the results connect smoothly to the spatial continuum limit, as it can be seen by varying $a$ for fixed volume $L = a N = 18.8/m$ (from dark to light blue
$\left\{am,N\right\} = \left\{0.94,20\right\},\left\{0.5875,32\right\},\left\{0.376,50\right\},\left\{0.2,94\right\}$; each with $\ell = 1500$). Cold-atom QED stands for the idealized parameters $am=0.05$ with $N = 376$ and $\ell = 5000$, for which the data are converged to $\ell\to \infty$ and the spatial continuum limit. 
}}
\end{figure}

\subsection{Numerical results for an alternative parameter set}
For comparison with Fig.~\ref{simulation2}, we show in Fig.~\ref{simulation1} the results for the second parameter set 
\begin{align}
&& am &= 0.94 \;, && e/m = 0.22 \; , && E_0/E_c = 3  \; , \nonumber
\end{align}
Qualitatively, the behaviour is very similar to what we found in Fig.~\ref{simulation2}, indicating that in this regime of weak coupling the slightly different values of $e/m$ do not qualitatively affect the physics of the Schwinger mechanism. However, due to the relatively large lattice spacing, the electric field and the fermion density start to deviate more strongly from the expected continuum behaviour at about $t \approx \SI{70}{\milli \second}$.

\begin{figure}
	\centering{\includegraphics[scale=0.24]{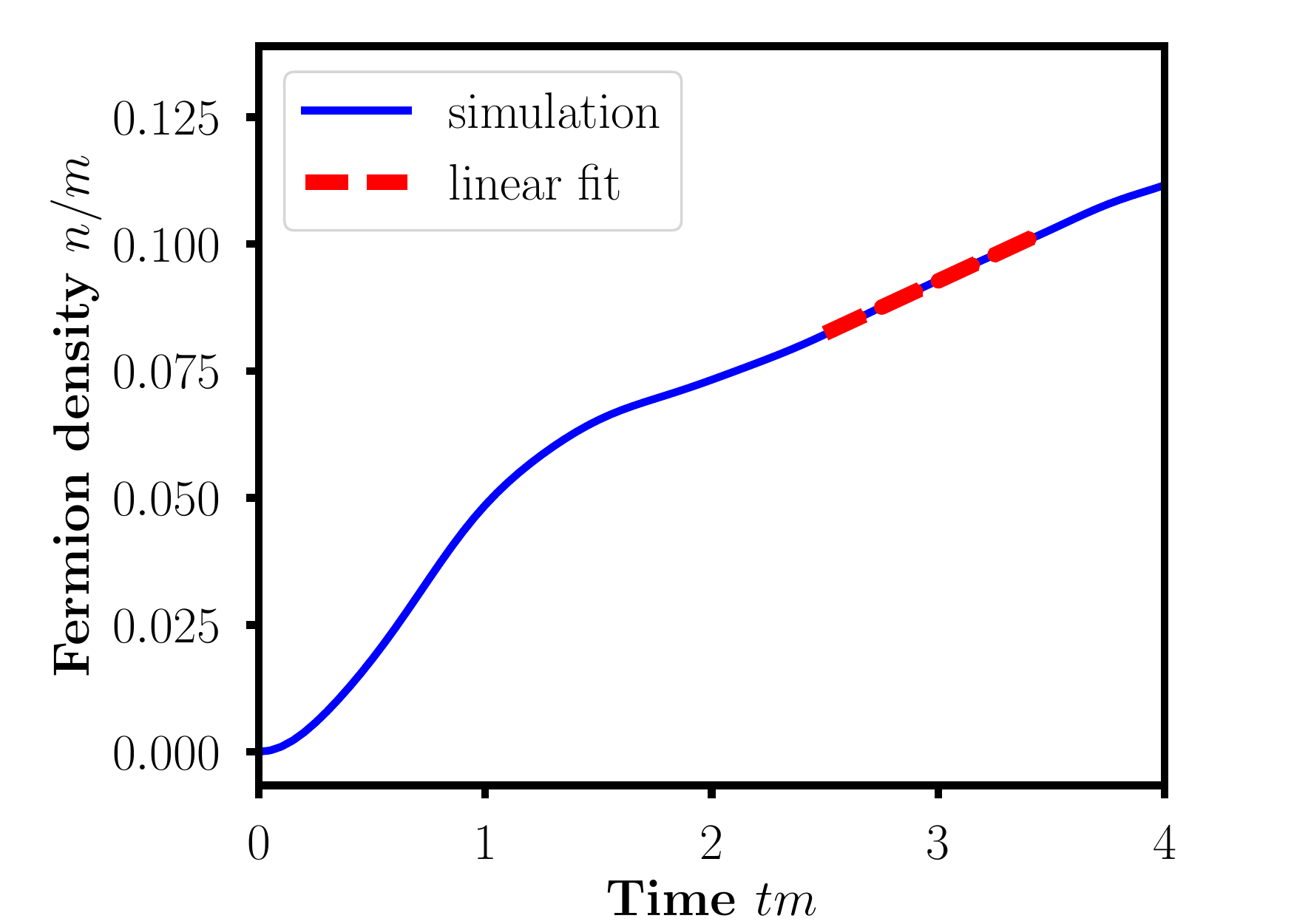}
		\includegraphics[scale=0.24]{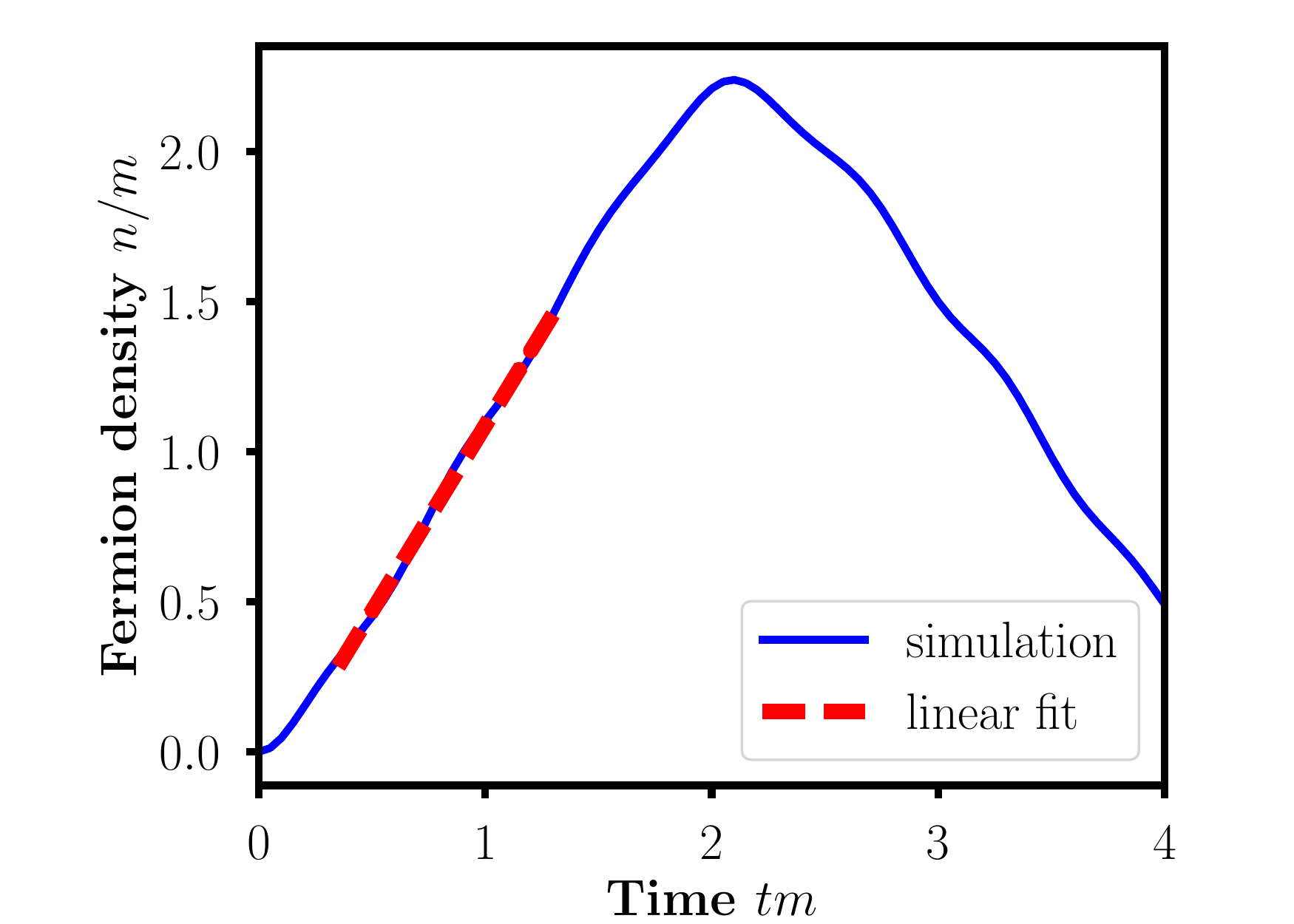}\caption{\label{fits}Examples of two linear fits underlying Fig.~\ref{rate_E}. Left: $E_0/E_c = 1$ with the fit in the linear regime after the initial quench. Right: $E_0/E_c = 7$ with the fit at early times before the occurrence of lattice artifacts. The time interval for the fit is fixed to the constant value of $\delta t=1/m$.}}
\end{figure}

\subsection{Details concerning Fig.~\ref{rate_E}, particle production rate}
We extract the particle production rate by fitting a linear function with an offset to the simulated particle number. The final and initial points for this fit have to be adjusted due to the following two reasons. First, the initial value problem considered here can be understood as a quench of the electric field. Therefore, the very early time dynamics is dominated by this quench and not by the Schwinger mechanism, which is a many-body phenomenon appearing in the long-time limit. Second, the simulation is limited by lattice artifacts as we already pointed out in appendix \ref{app:B}. The reason is that the produced particles are accelarated and invalidate the simulation as soon as they reach the boundary of the Brillouin zone. We empirically find that we can avoid both complications if we take $\delta t = 1/m$ as a constant window length for the fit and adjust the initial point accordingly. In Fig.~\ref{fits}, we show the fits for the largest and smallest values of $E_0$ that we considered.

In order to estimate the experimental error of the rescaled rate, we tentatively assume that the bosonic particle number imbalance $\langle b_n^\dagger b_n - d_n^\dagger d_n \rangle$ can be measured with an absolute accuracy of $10$ particles on each site. Then the electric field $E/E_c$ has an uncertainty of $\Delta \left(E/E_c\right) \approx 0.07$, see Eq.~\eqref{Efield}. We have further included the standard deviation of the fits in the calculation of the errors, although it is negligibly small compared to the uncertainty of the electric field. 
Though both errors are barely visible in total rate (inset of Fig.~\ref{rate_E}), they lead to significant uncertainties of the rescaled rate for larger fields (main panel of Fig.~\ref{rate_E}). 
Nevertheless, the non-perturbative behavior of the Schwinger rate can be clearly seen, and the qualitative agreement with the analytical predictions are satisfactory, in particular in light of the fact that the latter are derived for the limit of infinitely large times.

To close this section, let us comment on the experimental challenges to extract the rate. The total particle number corresponding to the density $n/m=1$ is $ n/m \times L = 9$,  thus on average $\mathcal{O}(10)$ respectively $\mathcal{O}(0.1)$ particles are produced over the whole lattice for the initial value $E= 7 E_c$ respectively $E= 1E_c$. 
Consequently, while the large-field case seems reasonably accessible, in the weak-field case the precise detection of the produced particles is very challenging with current technology. Moreover, the measurement of the rate requires an even higher accuracy as compared to the total particle number. This is more relevant for the weak-field regime, where the initial quench dynamics dominates the total particle number production at the short times accessible in the experiment.
Concerning the time-scales, we finally note that the fits for $E < 7 E_c$ require an observation time of up to $\sim\SI{400}{\milli \second}$.  
This may limit the observability of rates at small fields, since then three body losses become increasingly important, see Sec.~\ref{sec:experimentalLimitations}.

\begin{figure}\centering{
\includegraphics[scale=0.35]{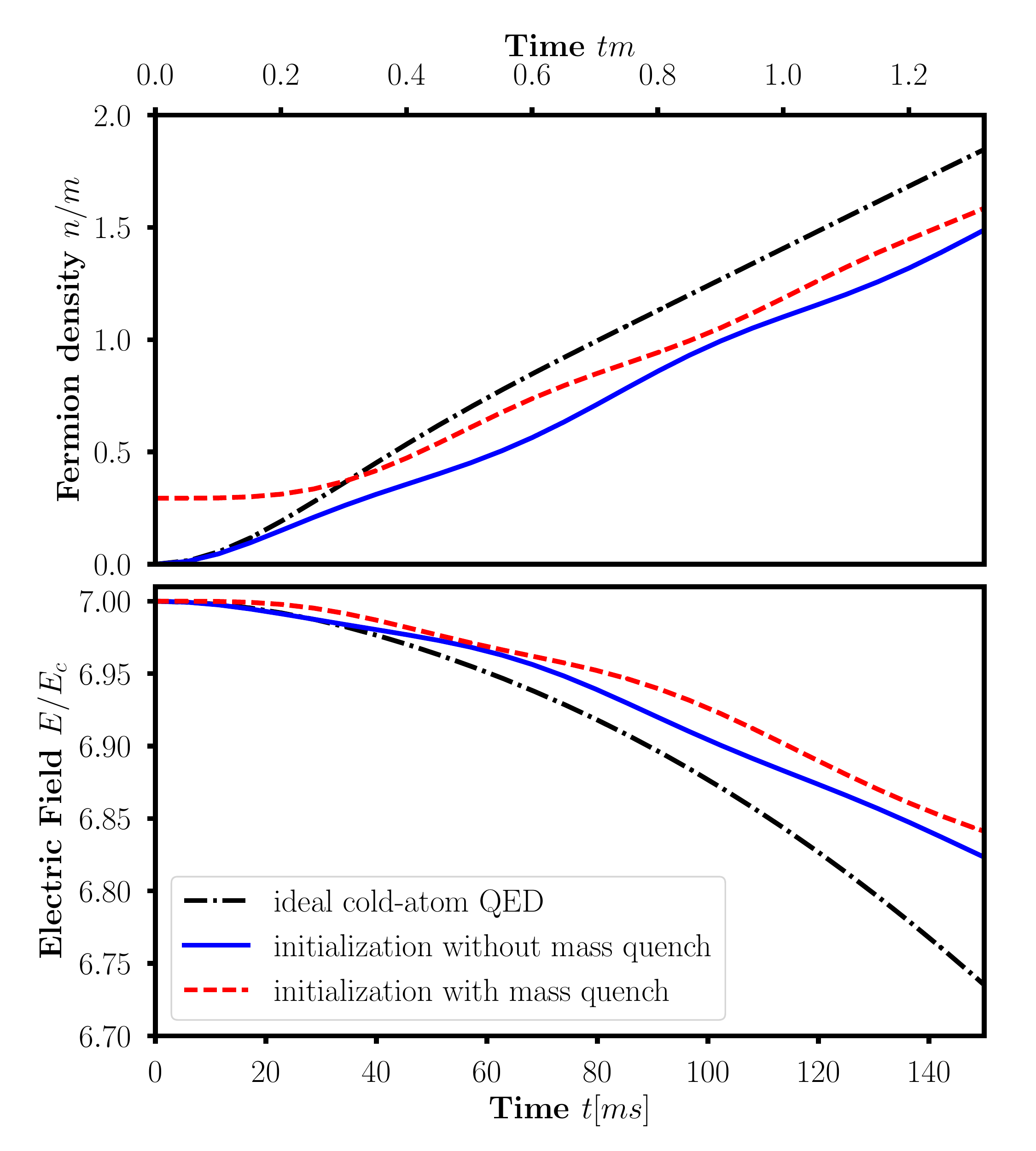}
\caption{\label{mass_quench} The mass-quenched simulation ($e/m = 0.45$, $am=0.45$, $N=20$, and $\ell = 1500$; red dashed line) shows a similiar growth of the fermion density as the simulation without an initial mass quench (same lattice parameters; blue solid line). The main qualitative difference for this parameter regime is the non-vanishing initial fermion density, which is due to the mass quench not realizing the vacuum of the free fermions. Both experimentally relevant simulations yield a growth rate comparable to the ideal continuum result (the limit of small $am$ as well as large $N$ and $\ell$; black dash-dotted). 
}}
\end{figure}

\subsection{Results for the initial infinite-mass vacuum}
In this subsection, we present results for the experimentally more feasible initialization of the free fermion vacuum for infinite mass $m$. Therefore we initialize the numerical simulation according to the fermion correlators \eqref{intial_fermion_corr} with $m \rightarrow \infty$ and solve the same equations of motions as before (with finite $m$). This corresponds to an additional quench of the fermion mass that is absent in the adiabatic preparation of the fermion vacuum. Figure \ref{mass_quench} compares the extracted particle number density and the electric field for this initial condition to the data already shown in the main part. Due to the additional mass quench, the initial state is not the true vacuum of free fermions and thus the system starts at non-vanishing fermion density. Even though the quantitative behaviour of the particle-production is quite different, the production \emph{rate} is very similar to the previously shown result. At the short times presented here, the electric field also shows qualitatively the same behaviour as for the simulations without mass quench. For comparison, we have included the simulations with mass quench in the analysis of the production rate of electron-positron pairs shown in figure \ref{rate_E}.

The consistence of the results leads us to conclude that the quantum simulator could also be benchmarked following the experimentally simpler initialization procedure, at least for the experimental and resulting lattice parameters that we have chosen in this work.

\appendix

\bibliographystyle{apsrev4-1}
\bibliography{0publications}

\end{document}